\documentclass[acmtocl,acmnow]{acmtrans2m}     

\usepackage[basic,small]{complexity} 
\usepackage{subfigure}
\usepackage{floatflt}
\usepackage{pslatex}
\usepackage[T1]{fontenc}
\usepackage{graphicx}
\usepackage{floatflt}
\usepackage{amsmath}
\usepackage{amsfonts}
\usepackage{amssymb}
\usepackage{pslatex}
\usepackage{color}
\usepackage{array}

\newtheorem{theorem}{Theorem}[section]

\newtheorem{corollary}[theorem]{Corollary}
\newtheorem{proposition}[theorem]{Proposition}
\newtheorem{lemma}[theorem]{Lemma}
\newdef{definition}[theorem]{Definition}
\newdef{remark}[theorem]{Remark}
\newdef{notation}[theorem]{Notation}

\newcommand{\Ende}{\qed}        

\mathchardef\ls="213C    
\mathchardef\gr="213E    

\newcommand{\Nat}{{\mathbb{N}}} 
\renewcommand{\epsilon}{\varepsilon} 
\newcommand{\sat}{\models}      
\newcommand{\impl}{\Rightarrow} 

\newcommand{\paccept}{{{\mathtt{accept}}}}
\newcommand{\pfail}{{{\mathtt{fail}}}}
\newcommand{\pretry}{{{\mathtt{retry}}}}
\newcommand{\psink}{{{\mathtt{sink}}}}
\newcommand{\psuccess}{{{\mathtt{success}}}}
\newcommand{\tin}{{{\mathtt{in}}}}
\newcommand{\tout}{{{\mathtt{out}}}}

\newcommand{\ttdol}{\texttt{\$}}

\newcommand{\A}{{\mathcal{A}}}  
\renewcommand{\c}{{c}} 
\renewcommand{\C}{{\mathsf{C}}} 
\renewcommand{\L}{{\mathcal{L}}} 
\newcommand{\N}{{\mathcal{N}}}  
\newcommand{\Pcal}{{{\mathcal{P}}}} 
\newcommand{\bfP}{{\textbf{P}}} 
\newcommand{\U}{{\mathcal{U}}}  
\newcommand{\V}{{\mathcal{V}}}  
\renewcommand{\W}{{\mathcal{W}}}        
\newcommand{\Until}{\mathrel{\mathit{Until}}}

\newcommand{\Acc}{{\mathit{Acc}}}
\newcommand{\Conf}{{\mathsf{Conf}}}
\newcommand{\false}{{\mathsf{false}}}
\newcommand{\Graph}{{\mathit{Graph}}}

\newcommand{\lost}{{\mathit{lost}}}
\newcommand{\MC}{{\mathit{MC}}} 
\newcommand{\LTS}{{\mathit{LTS}}} 
\newcommand{\MDP}{{\mathit{MDP}}} 
\newcommand{\Mess}{{\mathsf{M}}} 
\newcommand{\op}{{\mathit{op}}}
\newcommand{\Prom}{{\mathit{Prom}}}
\newcommand{\Safe}{{\mathit{Safe}}}
\newcommand{\true}{{\mathsf{true}}}

\newcommand{\overto}[1]{\xrightarrow{\!\!#1\!\!}}
\newcommand{\step}[1]{\overto{#1}} 

\newcommand{\egdef}{\stackrel{\mbox{\begin{scriptsize}def\end{scriptsize}}}{=}}

\title{Verifying nondeterministic probabilistic channel systems against $\omega$-regular linear-time properties}
            
\author{CHRISTEL BAIER \\ Universit{\"a}t Bonn, Institut f{\"u}r Informatik
  I \and NATHALIE BERTRAND and PHILIPPE SCHNOEBELEN \\ Lab.\ Specification
  \& Verification, CNRS \& ENS de Cachan}
            
\begin{abstract}
  Lossy channel systems (LCS's) are systems of finite state processes that
  communicate via unreliable unbounded fifo channels. We introduce NPLCS's,
  a variant of LCS's where message losses have a probabilistic behavior
  while the component processes behave nondeterministically, and study the
  decidability of qualitative verification problems for $\omega$-regular
  linear-time properties.

  We show that -- in contrast to finite-state Markov decision processes --
  the satisfaction relation for linear-time formulas depends on the type of
  schedulers that resolve the nondeterminism. While the qualitative model
  checking problems for the full class of history-dependent schedulers is
  undecidable, the same questions for finite-memory schedulers can be
  solved algorithmically. Additionally, some special kinds of reachability,
  or recurrent reachability, qualitative properties yield decidable
  verification problems for the full class of schedulers, which -- for this
  restricted class of problems -- are as powerful as finite-memory
  schedulers, or even a subclass of them.
\end{abstract}


            

\begin{document}

\begin{bottomstuff}
  Authors' adresses: C.~Baier, Universit{\"a}t Bonn, Institut f{\"u}r
  Informatik, R{\"o}merstr. 164, D-53117 Bonn, Germany.\newline N.~Bertrand
  and Ph.~Schnoebelen, Laboratoire Sp{\'e}cification et V{\'e}rification, ENS de Cachan, 61 av.\ Pdt Wilson, 94235 Cachan Cedex,
  France.\newline This research was supported by \emph{Pers{\'e}e}, a
  project of  the ACI S{\'e}curit{\'e} Informatique,  by \emph{PROBPOR}, a
  DFG-project, and by \emph{VOSS}, a DFG-NWO-project.
\end{bottomstuff}

\maketitle


\section{Introduction}

\begin{figure}[htbp]
\vspace{-.8em}
\centering
\includegraphics{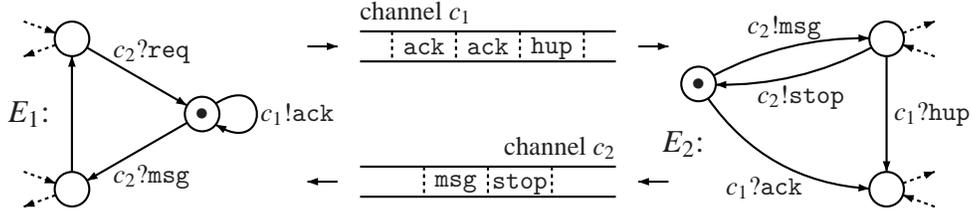}
\vspace{-.5em}
\caption{A channel system: $E_1$ and $E_2$ communicate through channels $c_1$ and $c_2$.}
\label{fig-exmpcs}
\end{figure}

Channel systems~\cite{brand83} are systems of finite-state components that
communicate via asynchronous unbounded fifo channels.
See Fig.~\ref{fig-exmpcs} for 
an example of a channel systems with two components $E_1$ and $E_2$
that communicate through
fifo channels $c_1$ and $c_2$.
\emph{Lossy channel systems}~\cite{finkel94,abdulla96b} are a special class
of channel systems where messages can be lost while they are in transit,
without any notification. Considering lossy systems is natural when modeling
fault-tolerant protocols where the communication channels are not supposed
to be reliable. Additionally, the lossiness assumption makes termination and
safety properties decidable~\cite{pachl87,finkel94,cece95,abdulla96b}.
Several important verification problems are undecidable for these
systems, including recurrent reachability, liveness properties,
boundedness, and all behavioral
equivalences~\cite{abdulla96c,Sch-tacs2001,Mayr-unreliable}.
Furthermore, the above-mentioned decidable problems cannot be solved
in primitive-recursive time~\cite{phs-IPL2002}.

\paragraph*{Verifying Liveness Properties}
Lossy channel systems are a convenient model for verifying safety
properties of asynchronous protocols, and such verifications can sometimes
be performed automatically~\cite{abdulla-forward-lcs}. However, they are not so
adequate for verifying liveness properties. A first difficulty here is
the undecidability of liveness properties.

A second difficulty is that the model itself is too pessimistic when
liveness is considered. Protocols that have to deal with unreliable
channels usually have some coping mechanisms combining resends and
acknowledgments. But, without any assumption limiting message losses, no
such mechanism can ensure that some communication will eventually be
initiated. The classical solution to this problem is to add some fairness
assumptions on the channel message losses, e.g., ``if infinitely many
messages are sent through the channels, infinitely many of them will not be
lost''. However, fairness assumptions in lossy channel systems make
decidability more elusive~\cite{abdulla96c,MS-mfcs2002}.

\paragraph*{Probabilistic Losses}
When modeling protocols, it is natural to see message losses as some kind
of faults having a probabilistic behavior. Following this idea,
Purushothaman Iyer and Narasimha~\citeNN{purush97} introduced the first
Markov chain model for lossy channel systems, where message losses (and
other choices) are probabilistic.
In this model, verification of qualitative properties is decidable when
message losses have a high probability~\cite{baier99} and undecidable
otherwise~\cite{abdulla2005}.
An improved model was later
introduced by~\citeN{ABRS-icomp} where the probability of
losses is modeled more faithfully and where qualitative verification (and
approximate quantitative verification~\cite{rabinovich2003}) is decidable
independently of the likelihood of message losses. See the survey
by~\citeN{Sch-voss} for more details.

These models are rather successful in bringing back decidability. However,
they assume that the system is \emph{fully probabilistic}, i.e., the choice
between different actions is made probabilistically. But when modeling
channel systems, \emph{nondeterminism} is an essential feature. It is used to model the
interleaved behavior of distributed components, to model an unknown
environment, to delay implementation choices at early stages of the design,
and to abstract away from complex control structures at later stages.

\paragraph*{Our Contribution}
We introduce \emph{Nondeterministic} Probabilistic Lossy Channel Systems (NPLCS), a new model where channel systems behave nondeterministically
while messages are lost probabilistically, and for which the operational
semantics is given via
\emph{infinite-state Markov decision processes}. For these NPLCS's, we study
the decidability of qualitative  $\omega$-regular linear-time properties. We focus here on ``control-based'' properties, i.e., temporal
formulas where the control locations of the given NPLCS serve as atomic propositions.

There are eight variants of the qualitative verification problem for a
given $\omega$-regular property $\varphi$ and a starting configuration $s$, that
arise from
\begin{itemize}
\item the four types of whether $\varphi$ should hold almost surely
  (that is, with probability 1), with positive probability, with zero
  probability or with probability less than 1
\item existential or universal quantification over
all schedulers, i.e., instances that resolve the nondeterministic choices.
\end{itemize}
By duality of existential and universal quantification, it suffices to
consider the four types of probabilistic satisfaction and one variant
of quantification (existential or universal). We  deal with the
case of existential quantification since it is technically more convenient.

Our main results can be summarized as follows.  First, we present
algorithms for reachability properties stating that a certain set of
locations will eventually be visited.  We then discuss repeated
reachability properties.  While repeated reachability problems with
the three probabilistic satisfaction relations ``almost surely'',
``with zero probability'' and ``with probability less than 1'' can be
solved algorithmically, the question whether a certain set of
locations can be visited infinitely often ``with positive probability''
under some scheduler is undecidable.  It appears that this is because
schedulers are very powerful (e.g., they need not be recursive). In
order to recover decidability without sacrificing too much of the
model, we advocate restricting oneself to finite-memory schedulers, and
show this restriction makes the qualitative model checking problem  against
$\omega$-regular properties decidable for NPLCS's.

This article is partly based on, and extends, material presented in \cite{BS03,BS04}.
However, an important difference with this earlier work is that the NPLCS model we
use does not require the presence of idling steps (see
Remark~\ref{rem-idling} below). This explains why some of the results
presented here differ from those in \cite{BS03,BS04}.

\paragraph*{Outline of the Article}
Section~\ref{sec-nplcs} introduces probabilistic lossy channel systems
and their operational
semantics. Section~\ref{sec-safe-prom} establishes some
fundamentals properties, leading to algorithms for reachability and
repeated reachability problems (in section~\ref{sec-decid}).
Section~\ref{sec-undec} shows that some repeated reachability problems are
undecidable and contains other lower-bound results.
Section~\ref{sec-finite-mem} shows decidability 
for problems where attention is restricted to finite-memory schedulers,
and section~\ref{sec-omegareg} shows how positive results for Streett properties
generalize to arbitrary $\omega$-regular properties.
Finally, section~\ref{sec-conclusion} concludes the article.


\section{Nondeterministic probabilistic channel systems}
\label{sec-nplcs}

\paragraph*{\textbf{\emph{Lossy channel systems}}}
A lossy channel system (a LCS) is a tuple $\L=
(Q,\C,\Mess,\Delta)$ consisting of a finite set $Q=\{p,q,\ldots\}$ of
control \emph{locations} (also called \emph{control states}), a finite set
$\C=\{c,\ldots\}$ of \emph{channels}, a finite \emph{message alphabet}
$\Mess=\{m,\ldots\}$ and a finite set $\Delta=\{\delta,\ldots\}$ of
\emph{transition rules}. Each transition rule has the form $q
\step{\op} p$ where $\op$ is an \emph{operation} of the form
\begin{itemize}
\item
\ $\c!m$ (sending message $m$ along channel $\c$),
\item
\ $\c?m$ (receiving message $m$ from channel $\c$),
\item
\ $\surd$ (an internal action to some process, no I/O-operation).
\end{itemize}
The \emph{control graph} of $\L$ is the directed graph having the locations
of $\L$ as its nodes and rules from $\Delta$ for its edges. It is denoted
with $\Graph(Q)$, and more generally $\Graph(A)$ for $A \subseteq Q$ denote
the control graph restricted to locations in $A$.
\\

Our introductory example in Fig.~\ref{fig-exmpcs} is turned into a LCS by
replacing the two finite-state communicating agents $E_1$ and $E_2$ by the
single control automaton one obtains with the asynchronous product
$E_1\times E_2$.
\\

\emph{Operational Semantics.} Let $\L= (Q,\C,\Mess,\Delta)$ be a LCS.
A \emph{configuration}, also called \emph{global state}, is a pair
$(q,w)$ where $q \in Q$ is a location and $w : \C \to \Mess^*$ is a
channel valuation that associates with any channel its content (a
sequence of messages).  We write ${\Mess^*}^\C$ for the set of all
channel valuations, or just $\Mess^*$ when $|\C|=1$.  The set $Q\times
{\Mess^*}^\C$ of all configurations is denoted by $\Conf$. With abuse
of notations, we shall use the symbol $\epsilon$ for both the empty
word and the channel valuation where all channels are empty. If
$s=(q,w)$ is a configuration then we write $|s|$ for the total number
of messages in $s$, i.e., $|s|=|w|= \sum_{\c \in \C} |w(\c)|$.

We say that a transition rule $\delta = q \step{\op} p$ is \emph{enabled} in
configuration $s = (r,w)$ iff
\begin{enumerate}
\item the current location is $q$, i.e., $r=q$, and
\item performing $\op$ is possible. This may depend on the channels
   contents: sending and internal actions are always enabled, while a
   receiving $\c?m$ is only possible if the current content of channel $\c$
   starts with the message $m$, i.e., if the word $w(\c)$ belongs to $m
   \Mess^*$.
\end{enumerate}
For $s$ a configuration, we write $\Delta(s)$ for the set of transition
rules that are enabled in $s$.

When $\delta=p \step{\op} q$ is enabled in $s=(q,w)$, firing $\delta$
yields a configuration $s'=(p,\op(w))$ where $\op(w)$ denotes the new
contents after executing $\op$:
\begin{itemize}
\item if $\op = \surd$, then $\op(w) = w$,
\item if $\op = \c!m$, then $\op(w)(\c) = w(\c)m$, and $\op(w)(\c') =
w(\c')$ for $\c \neq \c'$,
\item if $\op = \c?m$ (and then $w(\c)$ is some $m\mu$ since $\delta$ was
enabled), then $\op(w)(\c) = \mu$, and $\op(w)(\c') = w(\c')$ for $\c \neq
\c'$.
\end{itemize}
We write $s \step{\delta}_{\textrm{perf}} s'$ when $s'$ is obtained by
firing $\delta$ in $s$. The ``\textrm{perf}'' subscript stresses that the
step is perfect: no messages are lost.

However, in lossy systems, arbitrary messages can be lost. This is
formalized with the help of the subword ordering: we write $\mu \sqsubseteq
\mu'$ when $\mu$ is a subword of $\mu'$, i.e., $\mu$ can be obtained by
removing any number of messages from $\mu'$, and we extend this to
configurations, writing $(q,w)\sqsubseteq (q',w')$ when $q=q'$ and
$w(\c)\sqsubseteq w'(\c)$ for all $c\in\C$. By Higman's Lemma,
$\sqsubseteq$ is a well-quasi-order between configurations
of $\L$~\cite{abdulla2000c,finkel-wsts}. 

Now, we define lossy steps by letting $s \step{\delta}s''$ whenever there
is a perfect step $s \step{\delta}_{\textrm{perf}} s'$ such that $s''
\sqsubseteq s'$.\footnote{
   Note that, with this definition, message losses can only occur after
   perfect steps (thus, not in the initial configuration). This is usual
   for probabilistic models of LCS's, while nondeterministic models of
   LCS's usually allow losses both before and after perfect steps. In each
   setting, the chosen convention is the one that is technically smoother,
   and there are no real semantic differences between the two.
} This gives rise to a labeled transition system $\LTS_\L \egdef
(\Conf,\Delta,\to)$. Here the set $\Delta$ of transition rules serves as
action alphabet.

\begin{remark}
\label{remark:no-terminal-states}  
In the following we only consider LCS's where, for any location $q\in Q$,
$\Delta$ contains at least one rule $q \step{\op} p$ where $\op$ is not a
receive operation. This ensures that $\LTS_\L$ has no terminal
configuration, where no rules are enabled.
\Ende
\end{remark}


\begin{notation}[(Arrow-notations)]
  {\rm Let $s$, $t \in \Conf$ be configurations.  We write $s \to t$
    if $s \step{\delta} t$ for some $\delta$. As usual, $\step{+}$
    (resp.\ $\step{*}$) denotes the transitive (resp.\   reflexive and
    transitive) closure of $\to$. Let $\leadsto$ be $\to$, $\step{*}$
    or $\step{+}$. For $T \subseteq \Conf$, we write $s \leadsto T$
    when $s \leadsto t$ for some $t \in T$.  When $X \subseteq Q$ is a
    set of locations $s \leadsto X$ means that $s \leadsto (x,w)$ for
    some $x \in X$ (and for some $w$).

    We also use a special notation for \emph{constrained
      reachability}: $s \step{*}_{[X]} t$ means that there
    is a sequence of steps going from configuration $s$ to $t$ and
    \emph{visiting only locations from $X$}, including  at the two
    extremities $s$ and $t$. With $s \step{*}_{[X)} t$ we mean that the
    constraint does not apply to the last configuration. Hence  
    $s \step{*}_{[X)} s$ is always true, even with empty $X$.
    The following equivalence links the two notions:
    \begin{gather*}
        s \step{*}_{[X)} t
        \;\;\text{iff}\;\;
        \Bigl[s=t \;\text{ or }\;
        \exists s'\bigl(s \step{*}_{[X]} s' \text{ and } s'\step{}t\bigr)\Bigr].
    \Ende
    \end{gather*}
}
\end{notation}
We recall that in LCS's the following constrained reachability questions:
``given $s,t$ configurations, $X \subseteq Q$ and $\leadsto \in
\{\to,\step{*},\step{+} \}$ does $s \leadsto_{[X]} t$ (or $s \leadsto_{[X)}
t$)?'' are decidable \cite{abdulla96b,phs-IPL2002}.

\paragraph*{\textbf{\emph{The MDP-semantics}}}
Following Bertrand and Schnoebelen~\citeNN{BS03,BS04}, we define the
operational behavior of a LCS by an infinite-state Markov decision process.
A NPLCS\footnote{The starting letter ``N'' in NPLCS serves to indicate that
we deal with a semantic model where nondeterminism and probabilities
coexist, and thus, to distinguish our approach from interpretations of
probabilistic lossy channel systems by Markov chains.} $\N = (\L,\tau)$
consists of a LCS $\L$ and a \emph{fault rate} $\tau \in (0,1)$ that
specifies the probability that a given message stored in one of the message
queues is lost during a step. In the sequel, for $w,w'\in{\Mess^*}^\C$, 
we let $\bfP_\lost(w,w')$ denote the probability
that channels containing $w$ change to $w'$ within a  single step as a
result of message losses. This requires losing
$|w|-|w'|$ message at the right places.
Formally, we let
\begin{gather}
\label{proba-lost}
\bfP_\lost(w,w')
\egdef
\tau^{|w|-|w'|} \cdot (1-\tau)^{|w'|} \cdot \binom{w}{w'}
\end{gather}
where the combinatorial coefficient $\binom{w}{w'}$, is the number of
different embeddings of $w'$ in $w$. 
For instance, in the case where $w = aaba$, one has
\begin{center}
$\binom{aaba}{a} = \binom{aaba}{aa} = 3$, \ \
$\binom{aaba}{aba} = \binom{aaba}{ab} = 2$,
\ \ $\binom{aaba}{w'} = 1 \text{ \ if \ } 
   w'\in \{\varepsilon,b,aaa,aab,ba,aaba\}$ 
\end{center}
and $\binom{aaba}{w'} =0$ in all other cases. Note that, e.g., $w'= aa$ can
be obtained from $w = aaba$ in three different ways (by removing the $b$
and either the first, second or third $a$), while $w'= ba$ is obtained from
$w$ in a unique way (by removing the first two $a$'s).
See~\cite{ABRS-icomp} for more details. Here, it is enough to know that $\binom{w}{w'}\neq 0 \text{ iff } w'\sqsubseteq w$ and that the
probabilities add up to one: for all $w$, $\sum_{w'} \bfP_\lost(w,w')=1$.

The Markov decision process associated with $\N$ is
$\MDP_\N \egdef (\Conf,\Delta,\bfP_\N)$.
The stepwise probabilistic behavior is formalized by a
three-dimensional transition probability matrix $\bfP_\N : \Conf
\times \Delta \times \Conf \to [0,1]$.  For a given configuration $s$
and a transition rule $\delta$ that is enabled in $s$,
$\bfP_\N(s,\delta,\cdot)$ is a distribution over the states in
$\MDP_\N$, while $\bfP_\N(s,\delta,\cdot)=0$ for any transition rule
$\delta$ that is not enabled in $s$.  The intuitive meaning of
$\bfP_\N(s,\delta,t) = \lambda \gr 0$ is that with probability
$\lambda$, the system moves from configuration $s$ to configuration
$t$ when $\delta$ is the chosen transition rule in $s$.  Formally, if
$s = (q,w)$, $t = (p,w')$, and $\delta = q \step{\op} p$ is enabled in
$s$, then
\begin{gather}
\label{eq-Plost}
\bfP_\N(s,\delta,t) \egdef \bfP_\lost(\op(w),w').
\end{gather}

\begin{figure}[htbp]
\centering
\includegraphics{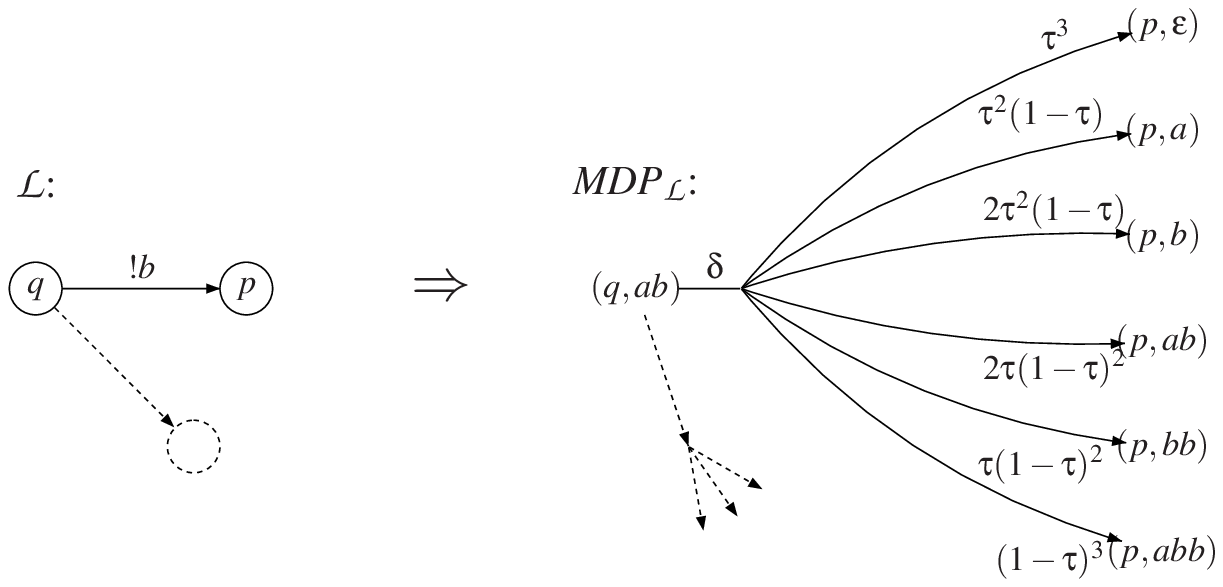}
\caption{From a LCS $\L$ to $MPD_\L$}
\label{fig-example}
\end{figure}

See Fig.~\ref{fig-example} for an example where $s=(q,ab)$ and $\delta= q\step{!b} p$.

A consequence of \eqref{proba-lost} and \eqref{eq-Plost} is that the
labeled transition system underlying $\MDP_\L$ is exactly $\LTS_\L$.
Hence any path in $\MDP_\L$ is also a path in $\LTS_\L$ and the fact
that $\LTS_\L$ had no terminal configuration implies that there is no
terminal state in $\MDP_\L$.



\begin{remark}[(The idling MDP semantics)]
\label{rem-idling}
The above definition of the MDP semantics for an NPLCS differs from the
approach of Bertrand and Schnoebelen~\citeNN{BS03,BS04} where each location
$q$ is assumed to be equipped with an implicit \emph{idling} transition
rule $q \step{\surd} q$. This idling MDP semantics allows simplifications
in algorithms, but it does not respect enough the intended liveness of
channel systems (e.g., inevitability becomes trivial) and we do not adopt
it here. Observe that the new approach is more general since idling rules
are allowed at any location in $\L$.
\Ende
\end{remark}

\paragraph*{\textbf{\emph{Schedulers (finite-memory, memoryless, blind and almost blind)}}}
Before one may speak of the probabilities of certain events in an MDP,
the nondeterminism has to be resolved by means of a scheduler, also
often called adversary, policy or strategy. We will use the word
``scheduler'' for a \emph{history-dependent deterministic scheduler}
in the classification of \citeN{Puterman94}. Formally, a scheduler for
$\N$ is a mapping $\U$ that assigns to any finite path $\pi$ in $\N$ a
transition rule $\delta \in \Delta$ that is enabled in the last state
of $\pi$.\footnote{As stated in Remark
  \ref{remark:no-terminal-states}, we make the assumption that any
  configuration has at least one enabled transition rule.}
Intuitively, the given path $\pi$ specifies the history of the system,
and $\U(\pi)$ is the rule that $\U$ chooses to fire next.

A scheduler $\U$ only gives rise to certain paths in the MDP: we say
$\pi=s_1\to s_2\to\cdots$ is \emph{compatible with $\U$} or, shortly, is a
\emph{$\U$-path}, if $\bfP_\N(s_n,\delta_n,s_{n+1}) \gr 0$ for all $n\geq
1$, where $\delta_n = \U(s_1 \to \cdots \to s_n)$ is the transition rule
chosen by $\U$ for the $n$-th prefix of $\pi$. In practice, it is only
relevant to define how $\U$ evaluates on $\U$-paths.

In general $\U$ can be any function and, e.g., it needs not be recursive.
It is often useful to consider restricted types of schedulers. In this
article, the two main types of restricted schedulers we use are
\emph{finite-memory schedulers}, that abstract the whole history into some
finite-state information, and \emph{blind schedulers}, that ignore the
contents of the channels.

Formally, a \emph{finite-memory} scheduler for $\N$ is a tuple $\U =
(U,D,\eta,u_0)$ where $U$ is a finite set of \emph{modes}, $u_0 \in U$ is
the \emph{starting mode}, $D : U \times \Conf \to \Delta$ is the
\emph{decision rule} which assigns to any pair $(u,s)$ consisting of a mode
$u \in U$ and a configuration $s$ a transition rule $\delta \in \Delta(s)$,
and $\eta : U \times \Conf \to U$ is a \emph{next-mode function} which
describes the mode-changes of the scheduler. The modes can be used to store
some relevant information about the history.
In a natural way, a finite-memory scheduler can be viewed as a scheduler in
the general sense: given a finite path $\pi = s_0 \to s_1 \to \cdots \to
s_n$ in $\N$, it chooses $D(u,s_n)$ where $u = \eta(u_0,s_0 s_1 \ldots s_n)
= \eta(\ldots\eta(\eta (u_0,s_0),s_1),\ldots,s_n)$.

A scheduler $\U$ is called \emph{memoryless} if $\U$ is finite-memory with
a single mode. Thus, memoryless schedulers make the same decision for all
paths that end up in the same configuration. In this sense, they are not
history-dependent and can be defined more simply via mappings $\U : \Conf
\to \Delta$.

By a \emph{blind} scheduler, we mean a scheduler where the decisions
only depend on the \emph{locations} that have been passed, and not on
the channel contents. Hence a blind scheduler never selects a reading
transition rule.
Observe that, since the probabilistic choices only affect channel contents
(by message losses), all $\U$-paths generated by a blind $\U$ visit the
same locations in the same
order. More formally, with any initial locations $q_0$, a blind scheduler
can be seen as associating an infinite sequence $q_0\step{\op_1} q_1
\step{\op_2} q_2 \cdots$ of chained transition rules and the $\U$-paths are
exactly the paths of the form
$(q_0,w_0) \to (q_1,w_1) \to (q_2,w_2) \to \cdots$
with $w_i\sqsubseteq \op_i(w_{i-1})$ for all $i>0$.

A scheduler is called \emph{almost blind} if it almost surely eventually
behaves blindly. Formally, $\U$ is almost blind iff there exists a
scheduler $\W$ and a blind scheduler $\V$ such that for all configurations
$s$ and for almost all (see below) infinite $\U$-paths $\pi = s_1 \to s_2
\to \cdots$ with $s=s_1$, there exists an index $n \geq 0$ such that
\begin{itemize}
\item
$\U(s_1 \to \cdots \to s_i) = \W(s_1 \to \cdots \to s_i)$
for all indices $i \leq n$ and
\item
$\U(s_1 \to \cdots \to s_i) = \V(s_1 \to \cdots \to s_i)$
\  for all indices $i \gr n$.
\end{itemize}
Here and in the sequel, the formulation ``almost all paths have property
$x$'' means that the paths where property $x$ is violated are contained in
some measurable set of paths that has probability measure 0. 
The underlying
probability space is the standard one (briefly explained below).

\paragraph*{\textbf{\emph{Stochastic process}}}
Given an NPLCS $\N$ and a scheduler $\U$, the behavior of $\N$ under $\U$
can be formalized by an infinite-state Markov chain $\MC_\U$. For arbitrary
schedulers, the states of $\MC_\U$ are finite paths in $\N$. Intuitively,
such a finite path $\pi = s_1 \to \cdots \to s_{n-1} \to s_n$ represents
configuration $s_n$, while $s_1 \to \cdots \to s_{n-1}$ stand for the
history how configuration $s_n$ was reached.\footnote{One often uses
informal but convenient formulations such as ``scheduler $\U$ is in
configuration $s$'', which means that a state $\pi$ in the chain $\MC_\U$,
i.e., a finite path in $\N$, is reached where the last configuration is
$s$.}
If $\pi$ is a finite path ending in configuration $s$, and $\pi'=\pi\to t$
is $\pi$ followed by step $s\to t$, then the probability
$\bfP_\U(\pi,\pi')$ in $\MC_\U$ is defined with
$\bfP_\U(\pi,\pi')\egdef\bfP_\N(s,\U(\pi),t)$, according to the chosen rule
$\U(\pi)$. In all other cases $\bfP_\U(\pi,\pi')=0$.
We now may apply the standard machinery for Markov chains and define (for
fixed starting configuration $s$) a sigma-field on the set of infinite
paths starting in $s$ and a probability measure on it, see,
e.g.,~\cite{Kemeny-Snell,Puterman94,panangaden2001}. We shall write
$\Pr_\U\bigl(s\models\cdots\bigr)$ to denote the standard probability measure in
$\MC_\U$ with starting state $s$.

For $\U$ a finite-memory scheduler, we can think of the states in $\MC_\U$
as pairs $(u,s)$ consisting of a mode $u$ and a configuration $s$.  In
the sequel, we will write $s_u$ rather than $(u,s)$ as the intuitive
meaning of $(u,s)$ is ``configuration $s$ in mode $u$''.
For finite-memory schedulers the successor-states of $s_u$ and their
probabilities in $\MC_\U$ are given by the MDP for $\N$ in configuration
$s$ and the chosen transition rule for $s_u$. That is, if $\U$ is some
$(U,D,\eta,u_0)$, we have $\bfP_\U(s_u,t_{\eta(u,s)}) \egdef
\bfP_\N(s,D(u,s),t)$, and if $u'\neq\eta(u,s)$ then
$\bfP_\U(s_u,t_{u'})=0$.
In a similar way, we can think of the Markov chains for memoryless or blind
schedulers in a simpler way. For memoryless schedulers, the configurations
of $\N$ can be viewed as states in the Markov chain $\MC_\U$, while for
blind schedulers we may deal with finite words over $Q$ complemented with
some current channel contents.

\paragraph*{\textbf{\emph{LTL-notation}}}
Throughout the article, we assume familiarity with linear temporal logic
(LTL), see, e.g., \cite{Emerson-Handbook}. We use
simple LTL formulas to denote properties of paths in $\MDP_\L$.
Here configurations and locations serve as atomic propositions: for example
$\Box \Diamond s$ (resp.\ $\Box \Diamond x$) means that $s \in \Conf$
(resp.\ $x \in Q$) is visited infinitely many times along a path, and $x
\Until s$ means that the control state remains $x$ until $s$ is eventually
reached.
These notations extend to sets: $\Box \Diamond T$ and $\Box \Diamond A$ for
$T\subseteq \Conf$ and $A \subseteq Q$ with obvious meanings. For $A
\subseteq Q$, $A_\epsilon$ is the set $\{(q,\epsilon) : q \in A\}$ so
that $\Diamond Q_\epsilon$ means that eventually a configuration with
empty channels is reached.
It is well-known that for any scheduler $\U$, the set of paths starting in
some configuration $s$ and satisfying an LTL formula, or an
$\omega$-regular property, $\varphi$ is measurable
\cite{Vardi85,Courcoubetis95}. 
We write $\Pr_\U\bigl(s
\models \varphi\bigr)$ for this measure.



\paragraph*{\textbf{\emph{Finite attractor}}}
The crucial point for the algorithmic analysis of NPLCS is the fact
that almost surely, a configuration where all channels are empty will
be visited infinitely often.  If $\U$ is a scheduler and $T$ a set of
configurations then $T$ is called an attractor for $\U$ iff
$\Pr_\U\bigl( s \models \Box \Diamond T \bigr) =1$ for any
starting configuration $s$.

\begin{proposition}[(Finite-attractor property for arbitrary schedulers)]
For any scheduler $\U$,
the set $Q_\epsilon = \{ (q,\epsilon) : q \in Q\}$
is a finite attractor for $\U$.
\end{proposition}

That is, almost all paths in $\MC_\U$ visit $Q_\epsilon$ infinitely often,
independent on the starting state. We refer to \cite{BS03,BBS-attractor}
for formal proofs. An intuitive explanation of the result is that when the
channels contain $n$ messages, each step can only add at most one new
message (through a sending action) while on average $n\times \tau$ are
lost. Thus when $n$ is large, it tends to decrease and this suffices to
ensure that almost surely all messages will be lost.


\section{Safe sets and Promising sets}
\label{sec-safe-prom}

At many places, our arguments use the notion of ``safe sets'' and
``promising sets'' of locations. In this section we define these
notions, relate them to behavioral features, and explain how to
compute them.

\subsection{Safe sets}
\label{ssec-safe}

\begin{definition}
  Let $\L=(Q,\C,\Mess,\Delta)$ be a lossy channel system and $A
  \subseteq Q$ be a set of locations. We say that $X \subseteq Q$ is
  \emph{safe} for $A$ if $X \subseteq A$ and $(x,\epsilon) \to X$
  for all $x \in X$.
\end{definition}

Assume $A \subseteq Q$. It is easy to see that if $X$ and $Y$ are
both safe for $A$, then $X \cup Y$ is safe for $A$ too. 
The same holds for infinite unions. 
As a
consequence, the largest safe set for $A$ exists (union of all safe
sets); it is denoted by $\Safe(A)$, or $\Safe$ when there is no
ambiguity on $A$.

Observe that for any family $(A_i)_{i \in I}$ of sets of locations,
one has the following inclusions
\begin{alignat}{2}
  \Safe\Bigl(\bigcup_{i \in I} A_i\Bigr) \:\supseteq\: \bigcup_{i\in I} \Safe(A_i) &
  \qquad \Safe\Bigl(\bigcap_{i \in I} A_i\Bigr) \:\subseteq\: \bigcap_{i\in I}
  \Safe(A_i)
\end{alignat}
while the reverse inclusions do not hold in general.



$\Safe(A)$ can be computed in linear time: consider $\Graph(A)$ the
control graph restricted to locations of $A$. Remove from $\Graph(A)$
the edges that carry receiving operations ``$\c ?m$''. The nodes that
have no outgoing edges cannot be in $\Safe(A)$: remove them with
their incoming edges. This may create new nodes with no outgoing
edges that have to be removed iteratively. After each iteration, the
remaining nodes are a superset of $\Safe(A)$. When the process
eventually terminates, what remains is exactly $\Safe(A)$. Indeed the
remaining nodes form a safe set $X$: from every $x \in X$ there is an
outgoing edge $x \step{\op} y$ where $\op$ is not a receiving, hence
$(x,\epsilon) \step{\op} X$.

The following lemma justifies the terminology ``safe'' and will be
very useful in the sequel.

\begin{lemma}
\label{lem-safe1}
There exists a blind and memoryless scheduler $\U$ s.t.\ for all $x \in
\Safe(A)$ and all $w \in {\Mess^*}^\C$, $\Pr_\U\bigl((x,w) \models \Box
A\bigr) =1$.
\end{lemma}
\begin{proof}
  Let us describe the scheduler $\U$ satisfying $\Box A$ with
  probability 1. For each $x \in \Safe(A)$ fix a rule $\delta_x : x
  \step{\op} y$ enabled in $(x,\epsilon)$ and with $y \in
  \Safe(A)$. One such rule must exist by definition of $\Safe(A)$.
  Because $y$ is in $\Safe$, $\U$ can go on with $\delta_y$, etc...
  Note that the rules used by $\U$ do not depend on the channels
  contents but only on the locations: this scheduler $\U$ is
  memoryless and blind. The fact that $\U$ fulfills the requirement
  $\Pr_\U\bigl((x,\epsilon) \models \Box A\bigr) =1$ comes for free
  from the inclusion $\Safe(A) \subseteq A$.
\end{proof}

Conversely:

\begin{lemma}
\label{lem-safe2}
If $\Pr_\U\bigl((x,\epsilon) \models \Box A\bigr) =1$ for some
scheduler $\U$, then $x \in\Safe(A)$.
\end{lemma}

\begin{proof}
  Assume $\Pr_\U\bigl((x,\epsilon) \models \Box A\bigr) =1$. We
  define $Y$ to be the set of locations that can be visited along a
  $\U$-path: $Y=\{q \in Q \mid \exists w,\ \Pr_\U\bigl((x,\epsilon)
  \models \Diamond (q,w)\bigr) \gr 0\}$ and show that $Y$ is safe for
  $A$. We have $Y \subseteq A$ otherwise $\Pr_\U\bigl((x,\epsilon)
  \models \Box A\bigr)$ would be less than $1$.

  Moreover, if $\Pr_\U\bigl((x,\epsilon) \models \Diamond
  (q,w)\bigr) \gr 0$ for some $w$ then $\Pr_\U\bigl((x,\epsilon)
  \models \Diamond (q,\epsilon)\bigr) \gr 0$. This is trivial if
  $q=x$, and otherwise, losing all messages in the last step leads to
  $(q,\epsilon)$ instead of $(q,w)$. Hence there must be some rule
  enabled in $(q,\epsilon)$ that $\U$ picks to satisfy $\Box A$
  with probability one. Let $q \step{\op} y$ this rule. Then $y$ is in
  $Y$.

  The set $Y$ is safe for $A$ and $x \in Y$, hence $x \in\Safe(A)$.
\end{proof}

\subsection{Promising sets}
\label{ssec-prom}

\begin{definition}
  Let $\L=(Q,\C,\Mess,\Delta)$ be a lossy channel system and $A
  \subseteq Q$ be a set of locations. We say that $X \subseteq Q$ is
  \emph{promising} for $A$ if $(x,\epsilon) \step{*}_{[X)} A$ for
  all $x \in X$.
\end{definition}
As for safe sets, the largest promising set for $A$ (written
$\Prom(A)$ or $\Prom$) exists: it is the union of all promising sets
for $A$.

An important property is distributivity with respect to union:
\begin{lemma}[(See Appendix~\ref{appendix-prom})]
\label{lem-prom-union}
For any family $(A_i)_{i \in I}$ of sets of locations,
\[
       \Prom\Bigl(\bigcup_{i \in I} A_i\Bigr) \:=\: \bigcup_{i \in I} \Prom(A_i).
\]
\end{lemma}

With regards to intersection, the following clearly holds:
\begin{gather}
  \Prom\Bigl(\bigcap_{i \in I} A_i\Bigr) \:\subseteq\: \bigcap_{i \in I} \Prom(A_i)
\end{gather}
but the reverse inclusion does not hold in general.

The set $\Prom(A)$ can be computed for a given $A$ as a greatest fixed
point. Let $X_0=Q$ be the set of all locations and, for
$i=0,1,\ldots$, define $X_{i+1}$ as the set of locations $x \in X_i$
such that $(x,\epsilon) \step{*}_{[X_i)} A$. The $X_i$'s can be built
effectively because constrained reachability is decidable for LCS's
(as recalled in section~\ref{sec-nplcs}).  The sequence eventually
stabilizes since $X_0=Q$ is finite. When it does $X\egdef\lim_{i} X_i$
is promising for $A$.  Since each $X_i$ is a superset of $\Prom(A)$,
we end up with $X=\Prom(A)$.

Promising sets are linked to eventuality properties:

\begin{lemma}
\label{lem-prom1}
There exists a memoryless scheduler $\U$ s.t.\ for all $x \in \Prom(A)$
and all $w \in {\Mess^*}^\C$, $\Pr_\U\bigl((x,w) \models \Diamond A\bigr)
=1$.
\end{lemma}

\begin{proof}
  We first describe a \emph{finite-memory} scheduler $\U$ that
  achieves for any $x \in \Prom(A)$ and $w \in {\Mess^*}^\C$,
  $\Pr_\U\bigl((x,w) \models \Diamond A\bigr) =1$. Then we explain
  how a \emph{memoryless} scheduler can do the same thing.

  $\U$ has two types of modes, a normal mode for each $x\in\Prom(A)$, and a
  recovery mode. In normal mode and starting from $(x,\epsilon)$ for
  some $x \in \Prom(A)$, $\U$ picks the rule $\delta_1$ given by a fixed
  path $\pi_x$ of the form $(x,\epsilon) \step{\delta_1} (x_1,w_1)
  \step{\delta_2} \cdots \step{\delta_n} A$ witnessing $x\in\Prom(A)$. If
  after firing $\delta_1$ the next configuration is indeed $(x_1,w_1)$,
  $\U$ stays in normal mode and goes on with $\delta_2$, $\delta_3$, etc.,
  trying to follow $\pi_x$ until $A$ is reached. Whenever the probabilistic
  losses put it out of $\pi_x$, i.e., in some $(x_i,w'_i)$ with
  $w'_i\neq w_i$ (and $x_i\notin A$), $\U$ switches to recovery mode.

  In recovery mode and in some configuration $(x_i,w)$, $\U$ performs a
  rule enabled in $(x_i,\epsilon)$ and leading to a location $y \in
  \Prom(A)$ -- such a rule exists because $x_i \in \Prom(A)$, e.g., the
  first rule used in $\pi_{x_i}$. $\U$ goes on in recovery mode until all
  channels are empty. Note that in normal mode and in recovery mode all the
  visited locations are in $\Prom(A)$. Because of the finite-attractor
  property, with probability one some configuration $(y,\epsilon)$ is
  eventually visited and $\U$ switches back to normal mode for $y$.
  Therefore, and as long as $A$ is not visited, some $\pi_x$ path is tried
  and almost surely one of them will be eventually followed to the end.
  Hence $\Pr_\U\bigl((x,w) \models \Diamond A\bigr) =1$. Observe that $\U$
  does not depend on $x$ (nor on $w$) and is finite memory.

  We can even design a memoryless scheduler, the so-called \emph{stubborn}
  scheduler. For this, it is enough to ensure that the set of paths
  $(\pi_x)_{x\in\Prom(A)}$ on which $\U$ relies are such that every
  occurring configuration is followed by the same next configuration. That
  is, the paths may join and fuse, but they may not cross and diverge (nor
  loop back). This way, $\U$ can base its choices on the current
  configuration only. Whether it is in ``normal'' or ``recovery'' mode is
  now based on whether the current configuration occurs in the set of
  selected paths or not.
\end{proof}

\begin{lemma}
\label{lem-prom2}
If $\ \Pr_\U\bigl((x,\epsilon) \models \Diamond A\bigr) =1$ for
some scheduler $\U$ then $x \in \Prom(A)$.
\end{lemma}

\begin{proof}
  Let $\U$ be a scheduler such that $\Pr_\U\bigl((x,\epsilon) \models
  \Diamond A\bigr)=1$. Define $X =\{y \in Q \mid\Pr_\U\bigl((x,\epsilon)
  \models \neg A \Until y\bigr) \gr 0\}$ and observe that $x\in X$.

  We now show that $X$ is promising for $A$. Let $y \in X$, then
  $\Pr_\U\bigl((x,\epsilon) \models \neg A \Until (y,\epsilon)\bigr)
  \gr 0$: this is obvious for $y=x$ and, for $y\neq x$, the channel can be
  emptied in the last step of the path witnessing $\neg A \Until y$.
  Thus, and since $\Pr_\U\bigl((x,\epsilon) \models \Diamond A\bigr)
  =1$, there must be some path $(y,\epsilon) \step{*} (z,w)$ with $z\in
  A$. Moreover if $z$ is the first occurrence of $A$ along this path, we
  have $(y,\epsilon) \step{*}_{[X)} (z,w)$.

  Hence $X$ is promising for $A$, and $x \in X$, so $x \in \Prom(A)$.
\end{proof}


\section{Decidability results}
\label{sec-decid}

\subsection{Reachability properties}
\label{ssec-decid-reachability}

In this section we give decidability results for qualitative
reachability problems. The questions whether there exists a scheduler
such that eventuality properties of the form $\bigwedge_i \Diamond
A_i$ are satisfied with probability $=1$ (resp.\ $=0$, $\gr 0$, $\ls 1$)
are all decidable.

In all cases the problem reduces to several reachability questions in
ordinary lossy channel systems.

\begin{theorem}[(Generalized eventuality properties)]
\label{thm:eventually}
It is decidable whether for a given NPLCS $\N$, location $q$, sets
$A_1,\ldots,A_n$ of locations and reachability properties (a), (b),
(c) or (d) there exists a scheduler $\U$ satisfying
\begin{itemize}
\item [\emph{(a)}] $\Pr_\U\bigl( (q,\epsilon)\models
  \bigwedge\limits_{i=1}^n \Diamond A_i\bigr) \gr 0$, or
\item [\emph{(b)}] $\Pr_\U\bigl( (q,\epsilon) \models
  \bigwedge\limits_{i=1}^n \Diamond A_i\bigr) = 0$, or
\item [\emph{(c)}] $\Pr_\U\bigl( (q,\epsilon)\models
  \bigwedge\limits_{i=1}^n \Diamond A_i\bigr) \ls 1$, or
\item [\emph{(d)}] $\Pr_\U\bigl( (q,\epsilon) \models
  \bigwedge\limits_{i=1}^n \Diamond A_i\bigr) =1$.
\end{itemize}
Furthermore, the existence of a scheduler $\U$ satisfying (b) entails
the existence of a \emph{blind and memoryless} scheduler for (b). The
existence of a scheduler satisfying (c) entails the existence of an
\emph{almost blind and memoryless} scheduler for (c). The existence of
a scheduler satisfying (a) or (d) entails the existence of a
\emph{finite-memory} scheduler for (a) or (d).
\end{theorem}

The rest of this section consists in the proof of
Theorem~\ref{thm:eventually}. In this proof, we will successively show
the decidability of (a), (b), (c) and (d).
\\

\noindent
\textbf{ad (a)} of Theorem~\ref{thm:eventually}: $\Pr_\U\bigl(
(q,\epsilon)\models \bigwedge\limits_{i=1}^n \Diamond A_i\bigr) \gr
0$.

We first consider the case of a single eventuality property
$\Diamond A$. Obviously:
\begin{center}
\begin{tabular}{rcl}
    &       & $A$ is reachable from $(q,\epsilon)$
\\
iff &       & there exists a scheduler $\U$ with $\Pr_\U\bigl( (q,\epsilon)
             \models \Diamond A) \bigr) \gr 0$
\\
iff &       & there exists a memoryless scheduler $\U$ with $\Pr_\U\bigl(
             (q,\epsilon) \models \Diamond A) \bigr) \gr 0$.
\end{tabular}
\end{center}
Hence the problem reduces to a control-state reachability problem in $\LTS_\L$.

For several eventualities $A_1,\ldots,A_n$, one can reduce the problem
to the simpler case by building a product $\N \times \A$ of $\N$ with
a finite-state automaton $\A$ that records which $A_i$'s have been
visited so far. $\N \times \A$ has $2^n$ times the size of $\N$. The
existence of a memoryless scheduler for $\N \times \A$ directly
translates into the existence of a finite-memory scheduler for $\N$.

Observe that for eventuality properties of the
form $\exists \U\ \Pr_\U\bigl( (q,\epsilon) \models \Diamond A
\wedge \Diamond B \bigr) \gr 0$, memoryless schedulers are not sufficient as
the only possibility to satisfy both constraints $\Diamond A$ and
$\Diamond B$ might be to visit a certain configuration $s$ twice and
to choose different transition rules when visiting $s$ the first and
the second time.
\\

\noindent
\textbf{ad (b)} of Theorem~\ref{thm:eventually}: $\Pr_\U\bigl(
(q,\epsilon) \models \bigwedge\limits_{i=1}^n \Diamond A_i\bigr) =
0$.

We rewrite the question as the existence of $\U$ such that
$\Pr_\U\bigl((q,\epsilon) \models \bigvee\limits_{i=1}^n \Box \neg
A_i\bigr) = 1$, or equivalently, with $B_i \egdef \neg A_i$, such that
$\Pr_\U\bigl((q,\epsilon) \models \bigvee\limits_{i=1}^n \Box
B_i\bigr) = 1$.

The next lemma reduces this question to a simple safety problem.

\begin{lemma}
\label{lem-red-even=0}
There exists a scheduler $\U$ with $\Pr_\U\bigl((q,\epsilon)
\models \bigvee\limits_{i=1}^n \Box B_i\bigr) = 1$ if and only if
there exists a blind and memoryless scheduler $\U$ with
$\Pr_\U\bigl((q,\epsilon) \models\Box B_i\bigr) = 1$ for some $i$,
$1 \leq i \leq n$.
\end{lemma}

\begin{proof}
  $(\Longleftarrow)$: is obvious.

  $(\Longrightarrow)$: We assume that $\Pr_\U\bigl((q,\epsilon)
  \models \bigvee\limits_{i=1}^n \Box B_i\bigr) = 1$.

  For all $I \subseteq \{1,\ldots,n\}$, $I \neq \emptyset$, let $X_I$
  be the set of all locations $x$ such that there exists a finite
  $\U$-path $\pi$ of the form $(q,\epsilon)=(x_0,w_0) \to (x_1,w_1)
  \to \cdots \to (x_m,w_m)=(x,w_m)$ satisfying:
\[
         \{x_0,\ldots,x_m\} \subseteq B_i \textrm{ iff } i \in I.
\]
Hence a path such as $\pi$ above witnesses that $x_m$ belongs to $X_I$
for $I$ the set of all indices $i$ such that $\pi \models \Box B_i$.

Let $I_x \egdef \bigl\{i \in \{1,\ldots,n\} \mid x \in B_i\bigr\}$. By
assumption $I_q$ is not empty and $q \in X_{I_q}$.

We now show, for all $I\neq\emptyset$, that
\begin{gather}
\label{eqcrucial}
X_I \subseteq \Bigl\{x \in \bigcap_{i \in I} B_i \:\Bigl|\Bigr.\:
(x,\epsilon) \to X_J \textrm{ for some } \emptyset \neq J \subseteq
I\Bigr\}.
\end{gather}
This can be seen as follows. Let $x \in X_I$. Then, there is a finite
path as above.  But then also
\[
(q,\epsilon)=(x_0,w_0) \to (x_1,w_1) \to \cdots \to (x_{m-1},w_{m-1}) \to
\underbrace{(x_m,\epsilon)}_{=(x,\epsilon)}
\]
is a $\U$-path.  Let $x \step{\op} y$ be the transition rule taken
by $\U$ for this path.  Then, $(x,\epsilon) \to (y,\epsilon)$.
Hence, there is an infinite $\U$-path $\pi$ starting with the prefix
\[
(q,\epsilon)=(x_0,w_0) \to (x_1,w_1) \to \cdots \to
(x_{m-1},w_{m-1}) \to (x,\epsilon) \to (y,\epsilon).
\]
Let $J
\egdef I \cap I_y$. $J$ is not empty because $\pi \models \Box B_i$ for
some $1 \leq i \leq n$. Moreover $(q,\epsilon)=(x_0,w_0) \to
(x_1,w_1) \to \cdots \to (x_{m-1},w_{m-1}) \to (x,\epsilon) \to
(y,\epsilon)$ is a witness for $y \in X_J$. Hence $(x,\epsilon)
\to X_J$.


We now construct simultaneously an infinite sequence $x_0,x_1,\ldots$
of locations and an infinite sequence $I_0,I_1,\ldots$ of sets on
indices with $x_0=q$ and s.t.\ $x_k\in X_{I_k}$ for $k=0,1,\ldots$ We
start with $I_0\egdef I_q$. At step $k$, $x_k\in X_{I_k}$ and
\eqref{eqcrucial} entail the existence of a step $(x_k,\epsilon)
\step{}X_J$ with $J\subseteq I_k$. We let $x_{k+1}$ be the smallest
$x\in X_J$ that can be reached from $x_k$ (assuming $Q$ is totally
ordered in some way) and $I_{k+1}\egdef J$.  Observe that
$I_0\supseteq I_1\supseteq \cdots$ and that $I_\infty$
($\egdef\bigcap_{k=0,1,\ldots}I_k$) is not empty thanks to
\eqref{eqcrucial}.  Observe that a scheduler $\V$ that visits
$x_0,x_1,\ldots$, is blind, satisfies $\bigwedge_{i\in I_\infty}\Box
B_i$, and only needs finite-memory, e.g., recording the current $I_k$.
A \emph{memoryless} scheduler $\U$ can be obtained from $\V$ by always
picking, for a location $x$, the rule that $\V$ picks last if $x$ is
encountered several times in the sequence $x_0,x_1,\ldots$. $\U$
visits less locations than $\V$, hence satisfies more $\Box B_i$
properties.
\end{proof}

Now, combining Lemmas~\ref{lem-red-even=0},~\ref{lem-prom1}
and~\ref{lem-prom2}, one sees that there exists a scheduler $\U$ with
$\Pr_\U\bigl((q,\epsilon) \models \bigvee\limits_{i=1}^n \Box B_i\bigr) =
1$ iff $q \in \bigcup_{i=1}^n \Safe(B_i)$, which is decidable since the
$\Safe(B_i)$'s can be computed effectively (section~\ref{ssec-safe}). This
concludes the proof of Theorem~\ref{thm:eventually}~(b).
\\

\noindent
\textbf{ad (c)} of Theorem~\ref{thm:eventually}: $\Pr_\U\bigl(
(q,\epsilon)\models \bigwedge\limits_{i=1}^n \Diamond A_i\bigr) \ls
1$.

We first observe that
\begin{center}
\begin{tabular}{rcl}
    &    & $\Pr_\U\bigl((q,\epsilon)\models \bigwedge\limits_{i=1}^n
            \Diamond A_i\bigr) \ls 1$
\\
iff &    & $\Pr_\U\bigl((q,\epsilon)\models \bigvee\limits_{i=1}^n \Box \neg
            A_i\bigr) \gr 0$
\\
iff &    & $\Pr_\U\bigl((q,\epsilon)\models \Box \neg A_i\bigr) \gr 0$ for
            some $i \in \{1,\ldots,n\}$.
\end{tabular}
\end{center}
Thus, it suffices to explain how to check whether there exists a
scheduler $\U$ with
\[
       \Pr\nolimits_\U\bigl( (q,\epsilon)\models \Box B\bigr) \gr 0
\]
where $B$ is a given set of locations.

The following lemma reduces our problem to a decidable reachability
question in $\LTS_\L$ (see (c.3)).
\begin{lemma}
The following assertions are equivalent:
\begin{itemize}
\item [\emph{(c.1)}]
There exists a scheduler $\U$ such that $\Pr_\U\bigl(
(q,\epsilon)\models \Box B\bigr) \gr 0$.
\item [\emph{(c.2)}]
There exists an almost blind, memoryless scheduler $\U$
with $\Pr_\U\bigl( (q,\epsilon)\models \Box B\bigr) \gr 0$.
\item [\emph{(c.3)}]
$(q,\epsilon) \step{*}_{[B]} \Safe(B)$.
\end{itemize}
\end{lemma}

\begin{proof}
(c.2) $\Longrightarrow$ (c.1): is obvious.

(c.3) $\Longrightarrow$ (c.2):



Let $\pi$ be a path witnessing $(q,\epsilon) \step{*}_{[B]}
\Safe(B)$. A scheduler $\U$ that tries to follow this path reaches
$\Safe(B)$ with positive probability. If $\pi$ is simple (i.e., loop-free) $\U$ is memoryless.
Whenever $\Safe(B)$ is reached, it is sufficient that $\U$ behave as the blind scheduler
for safe sets (Lemma~\ref{lem-safe1}). The resulting scheduler
is almost blind, memoryless, and achieves $\Pr_\U\bigl(
(q,\epsilon) \models \Box B \bigr) \gr 0$.

(c.1) $\Longrightarrow$ (c.3): Let $\U$ be a scheduler such that
$\Pr_\U\bigl( (q,\epsilon)\models \Box B\bigr) \gr 0$. Let
\[
X = \Bigl\{ x \in Q \:\Bigl|\Bigr.\:
        \Pr\nolimits_\U\bigl( (q,\epsilon) \models
        \Box \Diamond (x,\epsilon) \wedge \Box B \bigr) \gr 0 \Bigr\}.
\]
  The finite-attractor property yields that $X \neq \emptyset$.
  Moreover, each configuration $(x,\epsilon)$ with $x \in X$ is
  reachable from $(q,\epsilon)$ via a $\U$-path where $\Box B$
  holds.  Hence, we have
\[
                     (q,\epsilon) \step{*}_{[B]} X.
\]
  We now show that $X$ is safe for $B$, which yields $X \subseteq
  \Safe(B)$, and hence (c.3).

Obviously $X \subseteq B$. Now let $x \in X$. There exists a
transition rule $\delta_x = x \step{\op} y$ such that
\[
\Pr\nolimits_\U\bigl( (q,\epsilon) \models \Box \Diamond
(x,\epsilon) \wedge \text{``$\delta_x$ is chosen infinitely often
  in $(x,\epsilon)$''} \wedge \Box B \bigr) \gr 0.
\]
Since
$\bfP_\N\bigl( (x,\epsilon),\delta_x,(y,\epsilon) \bigr) \gr 0$,
we get
\[
        \Pr\nolimits_\U\bigl( (q,\epsilon) \models \Box \Diamond
        (x,\epsilon) \wedge \text{``$\delta_x$ is chosen infinitely often
        in $(x,\epsilon)$''} \wedge \Box \Diamond (y,\epsilon) \wedge \Box
        B \bigr) \gr 0.
\]
Hence, $\Pr_\U\bigl( (q,\epsilon) \models \Box \Diamond
(y,\epsilon) \wedge \Box B \bigr) \gr 0$.  This yields $y \in X$.
We conclude that there is a transition $(x,\epsilon) \to X$. As
this is true for any $x \in X$, $X$ is safe for $B$.
\end{proof}

\noindent
\textbf{ad (d)} of Theorem~\ref{thm:eventually}: $\Pr_\U\bigl(
(q,\epsilon)\models \bigwedge\limits_{i=1}^n \Diamond A_i\bigr) =
1$.

The case where $n=1$ is equivalent, by Lemmas~\ref{lem-prom1}
and~\ref{lem-prom2}, to $q \in \Prom(A_1)$, a decidable question.
Lemma~\ref{lem-prom1} shows moreover that a memoryless $\U$ (the
stubborn scheduler) is sufficient.

We now consider the general case. With any $I
\subseteq \{1,\ldots,n\}$ we associate a set $X_I \subseteq Q$ of
locations defined inductively with:
\begin{xalignat*}{2}
  X_\emptyset & \egdef Q
&
  X_I & \egdef \bigcup_{i \in I} \Prom(A_i \cap X_{I \setminus \{i\}})
        \textrm{ for } I \neq \emptyset
\end{xalignat*}
By Lemma~\ref{lem-prom-union} $X_I = \Prom(\bigcup_{i \in I} A_i \cap
X_{I \setminus \{i\}})$.

\begin{lemma}
  For all $I \subseteq \{1,\ldots,n\}$ there exists a finite-memory
  scheduler $\U_I$ such that $\forall q \in X_I\ \forall w\
  \Pr_{\U_I}((q,w) \models \bigwedge_{i \in I} \Diamond A_i)=1$.
\end{lemma}
\begin{proof}
The proof is by induction on (the size of) $I$.

For $I=\emptyset$, $\bigwedge_{i \in I} \Diamond A_i$ always holds.

Let $\emptyset \subsetneq I \subseteq \{1,\ldots,n\}$. The definition
of $X_I$ entails that there exists a memoryless scheduler $\U$ (see
Lemma~\ref{lem-prom1}) such that\
\[
  \forall q \in X_I\ \forall w\ \Pr\nolimits_\U\Bigl((q,w) \models \Diamond
  \bigcup_{i\in I} \bigl(X_{I \setminus \{i\}} \cap A_i\bigr)\Bigr)=1
\]
We now derive $\U_I$ out of $\U$: $\U_I$ behaves as $\U$ until some
configuration $(y,v)$ with $y \in X_{I \setminus \{i\}} \cap A_i$ (for
some $i \in I$) is reached. From that point $\U_I$ switches mode and
behaves as $\U_{I \setminus \{i\}}$. By induction hypothesis
$\bigwedge_{i \in I \setminus \{i\}} \Diamond A_i$ will be satisfied
almost surely from $(y,v)$. Hence $\Pr_{\U_I}\bigl((q,w) \models
\bigwedge_{i \in I} A_i\bigr)=1$. $\U_I$ is finite memory, since it
has at most one mode for each $I \subseteq \{1,\ldots,n\}$.
\end{proof}

\begin{lemma}
  For all $I \subseteq \{1,\ldots,n\}$, if $\Pr_\U\bigl(
  (q,\epsilon) \models \bigwedge_{i \in I} \Diamond A_i \bigr)=1$
  for some $\U$, then $q \in X_I$.
\end{lemma}

\begin{proof}
  Here again the proof is by induction on $I$.

  The case $I=\emptyset$ is trivial since $X_\emptyset =Q$.

  Let $\emptyset \subsetneq I \subseteq \{1,\ldots,n\}$ and assume
  $\Pr_\U\bigl( (q,\epsilon) \models \bigwedge_{i \in I} \Diamond
  A_i \bigr)=1$. We define
\[
  Y \egdef \{x \in Q \mid \exists \textrm{ a $\U$-path } \pi_x:\
  (q,\epsilon) \step{*}_{[B)} (x,\epsilon)\}
\]
where $B \egdef Q \setminus \bigcup_{i \in I} A_i$ and show that $Y
\subseteq X_I$. For a fixed $x \in Y$, since $\pi_x$ is a $\U$-path,
from $(x,\epsilon)$ there must be a path visiting all the $A_i$'s
for $i \in I$. Consider one such path and let $y$ be the first
location belonging to some $A_i$ for $i \in I$. Then $\pi'_x \egdef \
(q,\epsilon) \step{*} (x,\epsilon) \step{*}_{[\bigcap_{i \in I}
  A_i)} (y,\epsilon) \in A_i$ is again a $\U$-path. From
$(y,\epsilon)$, all the $A_i$'s with $i \in I \setminus \{i\}$ have
to be visited with probability one. Let $\U_y$ be a ``suffix''
scheduler of $\U$ given by: $\U_y((y,\epsilon) \to \cdots) \egdef
\U(\pi'_x \to \cdots)$. From the assumption on $\U$ and the form of
$\pi'_x$ we deduce that $\Pr_{\U_y}\bigl((y,\epsilon) \models
\bigwedge_{i \in I} \Diamond A_i\bigr)=1$. By induction hypothesis, $y \in
X_{I \setminus \{i\}}$. Hence $(x,\epsilon) \step{*}_{[Y)}
(y,\epsilon)$ entails $(x,\epsilon) \step{*}_{[Y)} \bigcup_{i
  \in I} A_i \cap X_{I \setminus \{i\}}$. By definition of $\Prom$
(greatest fixed point), $Y \subseteq \Prom(\bigcup_{i \in I} A_i \cap
X_{I \setminus \{i\}}) = X_I$. As a consequence $q \in Y$ implies $q
\in X_I$.
\end{proof}

\begin{corollary}
The following assertions are equivalent:
\begin{itemize}
\item [\emph{(d.1)}]
There exists a scheduler $\U$ with $\Pr_\U\bigl(
  (q,\epsilon) \models \bigwedge\limits_{i=1}^n \Diamond A_i\bigr)
  =1$.
\item [\emph{(d.2)}]
There exists a finite-memory scheduler $\U$ with
  $\Pr_\U\bigl( (q,\epsilon) \models \bigwedge\limits_{i=1}^n
  \Diamond A_i\bigr) =1$.
\item [\emph{(d.3)}]
$q \in X_{\{1,\ldots,n\}}$.
\end{itemize}
\end{corollary}

Hence decidability of (d.3) (see section~\ref{ssec-prom}) entails
decidability of (d.1).

\subsection{Repeated reachability properties}
\label{ssec-decid-Buechi}

We now discuss the decidability of repeated reachability problems,
formalized by a B\"uchi condition $\Box \Diamond A$ (``visit
infinitely often locations in $A$'') or generalized B\"uchi conditions
that arise through the conjunction of several B\"uchi conditions.


In this subsection, we see that for generalized B\"uchi conditions and
for the three probabilistic satisfaction criteria ``almost surely'',
``with zero probability'' or ``with probability $\ls 1$'' the class of
finite-memory schedulers is as powerful as the full class of
(history-dependent) schedulers. Furthermore the corresponding problems
can all be solved algorithmically. When the fourth criterion ``with
probability $\gr 0$'' is considered, the problem is undecidable (see
section~\ref{sec-undec}).

\begin{theorem}[(Generalized B\"uchi)]
\label{thm:Buechi-decid}
It is decidable whether for a given NPLCS $\N$, location $q$, sets
$A_1,\ldots,A_n$ of locations and repeated reachability properties
(a), (b) or (c) there exists a scheduler $\U$ satisfying
\begin{itemize}
\item [(a)] $\Pr_\U\bigl( (q,\epsilon)\models
  \bigwedge\limits_{i=1}^n \Box \Diamond A_i\bigr) = 1$, or
\item [(b)] $\Pr_\U\bigl( (q,\epsilon) \models
  \bigwedge\limits_{i=1}^n \Box \Diamond A_i\bigr) = 0$, or
\item [(c)] $\Pr_\U\bigl( (q,\epsilon)\models
  \bigwedge\limits_{i=1}^n \Box \Diamond A_i\bigr) \ls 1$.
\end{itemize}
Moreover, if such a scheduler exists then there is also a
\emph{finite-memory} scheduler with the same property. In case (b),
the existence of a scheduler entails the existence of an
\emph{almost-blind and memoryless} scheduler. In case (c), the
existence of a scheduler entails the existence of an
\emph{almost-blind and finite-memory} scheduler.
\end{theorem}

As for Theorem~\ref{thm:eventually} we show the decidability of (a),
(b) and (c) in turn.
\\

\noindent
\textbf{ad (a)} of Theorem~\ref{thm:Buechi-decid}: $\Pr_\U\bigl(
(q,\epsilon)\models \bigwedge\limits_{i=1}^n \Box \Diamond
A_i\bigr) =1$.

We prove the equivalence of the following three statements:
\begin{itemize}
\item [(a.1)]
There exists a scheduler $\U$ such that
  $\Pr_\U\bigl((q,\epsilon) \models \bigwedge\limits_{i=1}^n \Box
  \Diamond A_i \bigr) =1$.
\item [(a.2)]
There exists a finite-memory scheduler $\U$ such that
  $\Pr_\U\bigl((q,\epsilon) \models \bigwedge\limits_{i=1}^n \Box
  \Diamond A_i \bigr) =1$.
\item [(a.3)]
$q \in \bigcap\limits_{i=1}^n \Safe(\Prom(A_i))$.
\end{itemize}

\begin{proof}
(a.2) $\Longrightarrow$ (a.1): is obvious.

(a.1) $\Longrightarrow$ (a.3): Let $\U$ be a scheduler as in (a.1).
Let $X$ be the set of all locations $x \in Q$ that are visited
with positive probability under $\U$ starting from state $(q,\epsilon)$.  That is,
\[
X \egdef \Bigl\{ x \in Q \:\Bigl|\Bigr.\:
        \Pr\nolimits_\U\bigl((q,\epsilon) \models \Diamond x \bigr) \gr 0 \Bigr\}.
\]
Let us show that $X \subseteq \bigcap_{i=1}^n
\Safe(\Prom(A_i))$.

Any finite $\U$-path $(q,\epsilon) \step{*} s$ can be
extended to an infinite $\U$-path where $\bigwedge_{i=1}^n \Box
\Diamond A_i$ holds (otherwise, $\bigwedge_{i=1}^n \Box \Diamond A_i$
could not hold almost surely). Hence, for all $x \in X$, there must exist some $\U$-path
\[
  \pi \egdef (q,\epsilon) \step{*} (x,\epsilon) \step{+} A_1
  \step{+} A_2 \cdots \step{+} A_n \step{+} A_1 \cdots
\]
These paths only visit locations in $X$, hence witness
$X\subseteq\Prom(A_i)$ for all $i$. In turn, they also witness that $X$ is
safe for the $\Prom(A_i)$'s, hence $X \subseteq \bigcap_{i=1}^n
\Safe(\Prom(A_i))$. One concludes by noting that $q \in X$.

(a.3) $\Longrightarrow$ (a.2): Let $Y \egdef \bigcap_{i=1}^{n}
\Safe(\Prom(A_i))$ and assume $q\in Y$. For each $x \in Y$ and
$i=1,\ldots,n$
we pick a simple (i.e., loop-free) path $\pi_{x,i}$ of the form
\[
                     (x,\epsilon) \step{+}_{[Y]} A_i.
\]
We design a finite-memory
scheduler that works with the modes $(x,i)$ where $x \in Y$ and $1
\leq i \leq n$, and recovery modes $i$ for $1 \leq i \leq n$.
Intuitively, in the modes $(\cdot,i)$ $\U$ tries to reach $A_i$, using
the stubborn scheduler for $A_i$ (see proof of Lemma~\ref{lem-prom1}).
As soon as $A_i$ is reached, $\U$ changes to the mode $(\cdot,i+1)$
and tries to reach $A_{i+1}$ (here and in the sequel, we identify mode
$(x,1)$ with $(x,n+1)$).
As before, in recovery mode $i$, $\U$ just waits until a configuration
with empty channel is reached, staying in $\Safe(\Prom(A_i))$ in the
meantime. When some $(y,\epsilon)$ is eventually reached (which
happens almost surely due to the finite-attractor property),$\U$ switches
back to mode $(y,i)$.
Hence, $\U$ will almost surely eventually reach $A_i$.
But then, $\U$ switches to the modes for index $i+1$ and the same
argument applies for the next goal states $A_{i+1}$.  This yields
$\Pr_\U\bigl((q,\epsilon) \models \bigwedge_i \Box \Diamond A_i
\bigr) =1$, and $\U$ is a finite-memory scheduler.
\end{proof}

Decidability of (a) follows from decidability of (a.3) which is
established in section~\ref{sec-safe-prom}.
\\

\noindent
\textbf{ad (b)} of Theorem~\ref{thm:Buechi-decid}: $\Pr_\U\bigl(
(q,\epsilon)\models \bigwedge\limits_{i=1}^n \Box \Diamond
A_i\bigr) =0$.

Clearly,
\[
  \Pr\nolimits_\U\bigl( (q,\epsilon) \models \bigwedge\limits_{i=1}^n \Box
  \Diamond A_i \bigr) =0
        \ \ \text{iff}\ \
  \Pr\nolimits_\U\bigl( (q,\epsilon)
  \models \bigvee\limits_{i=1}^n \Diamond \Box \neg A_i \bigr) =1.
\]
Letting $B_i \egdef \neg A_i$, it suffices to show that it is
decidable whether there exists a scheduler $\U$ with
\[
        \Pr\nolimits_\U\bigl((q,\epsilon) \models \bigvee\limits_{i=1}^n
        \Diamond \Box B_i \bigr) =1.
\]
We show the equivalence of the following statements:
\begin{itemize}
\item [(b.1)]
There is a scheduler $\U$ with
  $\Pr_\U\bigl((q,\epsilon) \models \bigvee\limits_{i=1}^n \Diamond
  \Box B_i \bigr) =1$.
\item [(b.2)]
There is a finite-memory scheduler $\U$ with
  $\Pr_\U\bigl((q,\epsilon) \models \bigvee\limits_{i=1}^n \Diamond
  \Box B_i \bigr) =1$.
\item [(b.3)]
There is a scheduler $\V$ with $\Pr_\V\bigl(
     (q,\epsilon) \models \Diamond \bigcup\limits_{i=1}^n
     \Safe(B_i) \bigr) =1$.
\end{itemize}

\begin{proof}

(b.2) $\Longrightarrow$ (b.1): is obvious.

(b.1) $\Longrightarrow$ (b.3): We assume that we are given a scheduler
$\U$ as in (b.1). Let $X_i$ be the set of locations $x$ with
$\Pr_\U\bigl( (q,\epsilon) \models \Box \Diamond (x,\epsilon)
\wedge \Diamond \Box B_i \bigr) \gr 0$.  We then have $X_i \subseteq
B_i$.  We now show that
\begin{itemize}
\item [(i)] $(x,\epsilon) \to X_i$ for any $x \in X_i$, and
\item [(ii)] $\Pr_\U\bigl( (q,\epsilon) \models \Diamond
  \bigcup\limits_{i=1}^{n} X_i\bigr) =1$.
\end{itemize}
Note that (i) yields $X_i \subseteq \Safe(B_i)$. But then (ii) yields
 (b.3).
\\

\noindent
\textbf{Proof of (i):}
Let $x \in X_i$.  There exists a transition rule $\delta = x
\step{\op} y$ which is enabled in $(x,\epsilon)$ and such that
\[
\Pr\nolimits_\U\bigl( (q,\epsilon) \models \Box \Diamond
\bigl((x,\epsilon) \wedge \text{``$\delta$ is chosen for
  $(x,\epsilon)$''} \bigr) \wedge \Diamond \Box B_i \bigr) \gr 0.
\]
If the transition rule $\delta$ is chosen infinitely often in
configuration $(x,\epsilon)$ then almost surely the step
$(x,\epsilon) \step{} (y,\epsilon)$ occurs infinitely often.
Hence,
$\Pr\nolimits_\U\bigl( (q,\epsilon) \models \Box \Diamond
(x,\epsilon) \wedge \Box \Diamond (y,\epsilon) \wedge \Diamond
\Box B_i \bigr) \gr 0$ and thus $y \in X_i$.
\\

\noindent
\textbf{Proof of (ii):}
By definition of $X_i$, $\Pr_\U\bigl( (q,\epsilon) \models \Diamond
\Box B_i \wedge \Box \Diamond (z,\epsilon) \bigr) =0$ for any $z
\notin X_i$.  Hence, since $\Pr_\U\bigl( (q,\epsilon) \models
\bigvee_{i=1}^n \Diamond \Box A_i \bigr) =1$, for each $z \notin X
\egdef \bigcup X_i$ necessarily $\Pr_\U\bigl( (q,\epsilon) \models
\Box\Diamond (z,\epsilon) \bigr) =0$.  Hence,
\[
  \Pr\nolimits_\U\bigl( (q,\epsilon) \models \bigvee\limits_{z \notin X}
  \Box \Diamond (z,\epsilon) \bigr) =0.
\]
Thus, the finite-attractor property yields $\Pr_\U\bigl(
(q,\epsilon) \models \bigvee\limits_{x \in X} \Box \Diamond
(x,\epsilon) \bigr) =1$. In particular,
\[
  \Pr\nolimits_\U\bigl( (q,\epsilon) \models \Diamond
  \!\bigcup\limits_{i=1}^n X_i \bigr) =1.
\]

(b.3) $\Longrightarrow$ (b.2): Let $\V$ be a scheduler as in (b.3).
By Lemma~\ref{lem-prom1}, we may assume that $\V$ is memoryless.  We
then define $\U$ as the scheduler that behaves as $\V$ until a
location in $\bigcup_i \Safe(B_i)$ is reached (this happens almost
surely). When a location $x \in \Safe(B_i)$ is reached (for some $i$),
$\U$ mimics the so-called ``safe'' scheduler (blind and memoryless)
described in section~\ref{ssec-safe} for safe sets, and fulfills
$\Box \Safe(B_i)$ from location $x$ onwards. Since $\Safe(B_i)
\subseteq B_i$ we obtain $\Pr_\U\bigl( (q,\epsilon) \models
\bigvee_{i=1}^n \Diamond \Box B_i \bigr)=1$.  Moreover, $\U$ is an
almost blind, memoryless scheduler.
\end{proof}

\noindent
\textbf{ad (c)} of Theorem~\ref{thm:Buechi-decid}: $\Pr_\U\bigl(
(q,\epsilon)\models \bigwedge\limits_{i=1}^n \Box \Diamond
A_i\bigr) \ls 1$.

We first observe that for any scheduler $\U$:
\begin{center}
\begin{tabular}{rcl}
    &   & $\Pr_\U\bigl((q,\epsilon) \models \bigwedge\limits_{i=1}^n \Box
           \Diamond A_i \bigr) \ls 1$
\\
iff &   & $\Pr_\U\bigl((q,\epsilon) \models \bigvee\limits_{i=1}^n \Diamond
           \Box \neg A_i \bigr) \gr 0$
\\
iff &   & $\Pr_\U\bigl((q,\epsilon) \models \Diamond \Box \neg A_i \bigr)
           \gr 0$ for some $i \in \{1,\ldots,n\}$.
\end{tabular}
\end{center}
Hence, it suffices to discuss the decidability of the question whether
for a given set $B \subseteq Q$ there is a scheduler $\U$ with
$\Pr_\U\bigl( (q,\epsilon) \models \Diamond \Box B\bigr) \gr 0$.

The following statements are equivalent:
\begin{itemize}
\item [(c.1)]
$\Pr_\U \bigl( (q,\epsilon) \models \Diamond \Box B
  \bigr) \gr 0$ for some $\U$.
\item [(c.2)]
$\Pr_\U \bigl( (q,\epsilon) \models \Diamond \Box B
  \bigr) \gr 0$ for some \emph{almost blind and finite-memory} $\U$.
\item [(c.3)]
$(q,\epsilon) \step{*} \Safe(B)$.
\end{itemize}

\begin{proof}

(c.2) $\Longrightarrow$ (c.1): is obvious.

(c.3) $\Longrightarrow$ (c.2): Assume $\Safe(B)$ is reachable from
$(q,\epsilon)$. Then, there is a finite simple (i.e., loop-free) path $\pi$ from
$(q,\epsilon)$ to $(x,\epsilon)$ for some $x \in \Safe(B)$. Let
$\U$ be an almost blind, memoryless scheduler which generates the
above path $\pi$ with positive probability and when/if $\Safe(B)$ is
reached, behaves as the safe scheduler for $B$. Clearly, $\U$ has the
desired property.

(c.1) $\Longrightarrow$ (c.3): Let $\U$ be a scheduler as in (c.1).
We define $X$ to be the set of locations $x \in Q$ such that
$\Pr\nolimits_\U\bigl( (q,\epsilon) \models \Box \Diamond
(x,\epsilon) \wedge \Diamond \Box B \bigr) \gr 0$. 
The finite-attractor property entails that $X$ is not empty. Furthermore
$X$ is
reachable from $(q,\epsilon)$.  A reasoning as in the proof of
(b.1) $\Longrightarrow$ (b.3) (see proof of (i)) shows that $X$ is
safe for $B$.
\end{proof}
The decidability of (c.3) entails that (c) is decidable.


\section{Hardness and undecidability results}
\label{sec-undec}

In this section we investigate the computational complexity of the problems
shown decidable in section~\ref{sec-decid}, and we prove undecidability for
the remaining problems. Technically, most results are hardness proofs and
the involved reductions make repeated use of the following ``cleaning''
gadget.

\subsection{Cleaning gadget}
\label{ssec-cleaning-gadget}

\begin{figure}[htbp]
\centering
\includegraphics{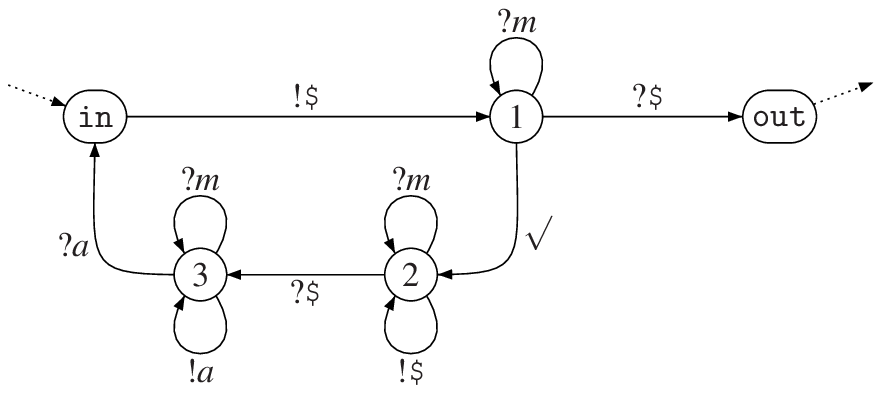}
\caption{Cleaning gadget, assuming $\ttdol\not\in\Mess$}
\label{fig-gadget}
\end{figure}

The cleaning gadget is the NPLCS shown in Fig.~\ref{fig-gadget}. It
can be part of a larger NPLCS where it serves to empty (``clean'') one
channel \emph{without introducing deadlocks}.
For a given message alphabet $\Mess=\{a,\ldots\}$, the system
described in Fig.~\ref{fig-gadget} uses one channel (left implicit)
and a new message symbol $\ttdol \notin \Mess$. Letter $a$ in
Fig.~\ref{fig-gadget} is a symbol from the original message alphabet
$\Mess$. Operations ``$?m$'' are used as a shorthand for all
$|\Mess|+1$ possible reading operations over the new message alphabet
$\Mess \cup \{\ttdol\}$.  The purpose of $\ttdol$ is to force the
channel to be emptied when moving from $\tin$ to $\tout$.

Let $T\subseteq\Conf$ be set of configurations described by the following
regular expression:
\[
  T = (\tin,\Mess^*) +(1,\Mess^*(\ttdol+\epsilon)) +(2,\Mess^*\ttdol^*)
  +(3,\ttdol^*a^*) +(\tout,\epsilon)
\]

\begin{lemma}
\label{lem-gadget}
The configurations reachable from $(\tin,\Mess^*)$ are exactly those
in $T$.
\end{lemma}
\begin{proof}[Sketch]
The left-to-right inclusion can be verified by showing that $T$ is an
invariant. For instance, from configurations $(\tin, \Mess^*)$ only the
configurations in $(1,\Mess^*(\ttdol + \epsilon))$ are reachable within one
step, while from $(2,\Mess^*\ttdol^*)$ only configurations in $(3,\ttdol^*)
+ (2,\Mess^*\ttdol^*)$ can be reached. And so on. The other inclusion is
easy to see.
\end{proof}
Constructions incorporating the gadget rely on the following property:
\begin{lemma}
\label{lem-gadget2}
  For any $w \in \Mess^*$:
\begin{itemize}
\item[\emph{(a)}] If $\U$ is a scheduler for the cleaning gadget and $v \neq\epsilon$ then $\Pr_\U\bigl( (\tin,w) \models \Diamond (\tout,v)
  \bigr) = 0$.
\item[\emph{(b)}] There is a (memoryless) scheduler $\U$ for the
  cleaning gadget with $\Pr_\U\bigl((\tin,w)\sat\Diamond
  (\tout,\epsilon) \bigr)=1$.
\end{itemize}
\end{lemma}

\begin{proof}
  (a) is immediate from Lemma~\ref{lem-gadget}. To prove (b),
  we describe a scheduler $\U$ with the desired property.  $\U$ starts
  from $(\tin,w)$, selects the $\tin\step{!\ttdol}1$ rule, aiming for configuration
  $(1,\ttdol)$ where $(\tout,\epsilon)$ can be reached. In case  a configuration
  $(1,v)$ with $v\neq\ttdol$  is reached, $\U$ moves from $1$ to $2$,
  goes back to $\tin$ and retry. This will eventually succeed with
  probability $1$.
\end{proof} 
Let us remark as an aside that, if one takes properties (a) and (b) above
as the specification of a cleaning gadget, then it can be proved that any
gadget necessarily uses ``new'' messages not from $\Mess$, like $\ttdol$ in
our construction.

\subsection{Complexity of decidable cases}
\label{ssec-complexity}

We consider the decidable cases given in section~\ref{sec-decid}. One
problem (reachability with zero probability) is in $\PTIME$, and even
$\NLOGSPACE$-complete, but all the others are non-primitive recursive,
as are most decidable problems for LCS's~\cite{phs-IPL2002}.

\begin{theorem}
  The problem, given NPLCS $\N$, location $q$ and set $A\subseteq Q$ of
  locations, whether there exists a scheduler $\U$ such that
  $\Pr_\U\bigl((q,\epsilon) \models \Box A \bigr) = 1$, is
  $\NLOGSPACE$-complete.
\end{theorem}

\begin{proof}[Sketch]
  Lemmas~\ref{lem-safe1} and~\ref{lem-safe2} show that the above
  problem is equivalent to a reachability question in some subgraph of
  the control graph of $\L$.
\end{proof}

\begin{theorem}
\label{non-prim}
The problem given a NPLCS $\N$, a location $q$ and a set of locations
$A$, whether there exists a scheduler $\U$ satisfying $(a.1)$ (or
$(a.2)$ ... or $(b.3)$), is not primitive recursive.
\[
\begin{array}{cclcccl}
\text{\emph{(a.1)}} &\;& \Pr_\U\bigl( (q,\epsilon)\models \Diamond A\bigr) \gr 0\text{, or}
&\;\;\;\;\;\;\;\;\;\;\;\;&
\text{\emph{(b.1)}} &\;& \Pr_\U\bigl( (q,\epsilon)\models \Box\Diamond A\bigr)= 0\text{, or}
\\[.3em]
\text{\emph{(a.2)}} && \Pr_\U\bigl( (q,\epsilon)\models \Diamond A\bigr) = 1\text{, or}
&& 
\text{\emph{(b.2)}} && \Pr_\U\bigl( (q,\epsilon)\models \Box\Diamond A\bigr) = 1\text{, or} 
\\[.3em]
\text{\emph{(a.3)}} && \Pr_\U\bigl( (q,\epsilon)\models \Diamond A\bigr) \ls 1\text{, or}
&&
\text{\emph{(b.3)}} && \Pr_\U\bigl( (q,\epsilon)\models \Box\Diamond A\bigr) \ls 1.
\end{array}
\]
\end{theorem}
In all six cases, the proof is by reducing from the control-state
reachability problem for (non-probabilistic) LCS's, known to be
non-primitive recursive \cite{phs-IPL2002}.
  
The case (a.1) is the easiest since, by Theorem~\ref{thm:eventually},
it is equivalent to the reachability of $A$ from $(q_0,\epsilon)$ in
the underlying LCS of $\N$.
  
\begin{figure}[htbp]
\centering
\includegraphics{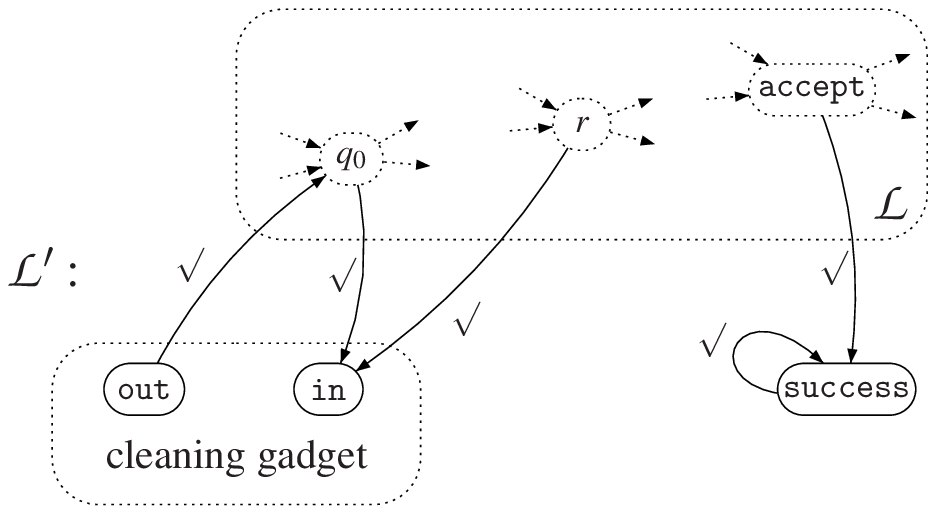}
\caption{The LCS $\L'$ associated with $\L$ in Lemma~\ref{lem-L'}}
\label{fig-reduc-3}
\end{figure}

For all the other cases, except (a.3), we use the reduction illustrated in
Fig.~\ref{fig-reduc-3}.  Let $\L$ be a LCS with only one channel and
two distinguished locations $q_0$ and $\paccept$. From $\L$ we build
another LCS $\L'$ and consider the NPLCS $\N=(\L',\tau)$ for any $\tau
\in (0,1)$. We now show that the control-state reachability problem in
$\L$ (i.e., is $\paccept$ reachable from $(q_0,\epsilon)$?) is
equivalent to particular instances of our probabilistic problems for
$\N$.

$\L'$ uses the cleaning gadget and has one further location:
$\psuccess$.  From every original location $r$ of $\L$, except
$\paccept$, $\L'$ has a $\surd$-transition to $\tin$, the input
location of the cleaning gadget. There is also a transition from
$\tout$ to $q_0$. From $\paccept$ there is a transition to $\psuccess$
and one can loop on this latter location.

The idea of this reduction is that, if $\paccept$ is reachable from
$q_0$ by some path $\pi$ in $\L$, then it is possible for a scheduler
to try and follow this path in $\L'$ and, in case probabilistic losses
do not comply with $\pi$, to retry as many times as it wants by
returning to $q_0$.  The cleaning gadget ensures that returning to
$q_0$ is with empty channel.  Note that the only way to visit
$\psuccess$ is to visit $\paccept$ first.  These general ideas are
formalized in the next lemma.

\begin{lemma}
\label{lem-L'}
In the LCS $\L'$, the following statements are equivalent:
\begin{itemize}
\item[\emph{(i)}] $(q_0,\epsilon) \step{*} \Prom(\{\psuccess\})$,
\item[\emph{(ii)}] $q_0 \in \Prom(\{\psuccess\})$,
\item[\emph{(iii)}] $(q_0,\epsilon) \step{*} \psuccess$,
\item[\emph{(iv)}] $(q_0,\epsilon) \step{*} \paccept$,
\item[\emph{(v)}] $(q_0,\epsilon) \step{*}_{[\L]} \paccept$,
\item[\emph{(vi)}] $q_0 \in \Safe(\Prom(\{\psuccess\}))$,
\item[\emph{(vii)}] $(q_0,\epsilon) \step{*} \Safe(\Prom(\{\psuccess\}))$.
\end{itemize}
\end{lemma}
Here ``$(q_0,\epsilon) \step{*}_{[\L]}\cdots$'' means that the path only
visits original locations from $\L$.
\begin{proof}
  (i) $\Longrightarrow$ (ii): Assume $(q_0,\epsilon) \step{*}
  \Prom(\{\psuccess\})$ and let $(q_0,\epsilon) \to (q_1,w_1) \to
  \cdots \to (q_m,w_m)$ with $q_m\in \Prom(\{\psuccess\})$ be a
  witness (simple) path.  From any $q_i\neq\psuccess$ along this path
  one may reach $(q_0,\epsilon)$ via the cleaning gadget. Hence
  $(q_i,\epsilon) \step{*} \Prom(\psuccess)$. All locations along the
  path from $(q_0,\epsilon)$ to $\Prom(\{\psuccess\})$ satisfies this
  property, hence we have $q_0 \in \Prom(\{\psuccess\})$.
  
  (ii) $\Longrightarrow$ (iii): by definition of $\Prom(.)$.
  
  (iii) $\Longrightarrow$ (iv): obvious.

  (iv) $\Longrightarrow$ (v): Assume $\pi$ is a path from $(q_0,\epsilon)$
  to $\paccept$. If this path steps out of $\L$ then it can only go to the
  cleaning gadget. From there the only exit back to $\L$ is via
  $(q_0,\epsilon)$ (Lemma~\ref{lem-gadget2}.(a)), looping back to a
  previously visited configuration. Thus if $\pi$ is a simple path, it
  stays inside $\L$.
  
  (v) $\Longrightarrow$ (vi): suppose $(q_0,\epsilon) \step{*}_{[\L]}
  \paccept$. Then $(q_0,\epsilon) \step{*} \psuccess$ and
  $s\step{*}\psuccess$ for all configurations of $\L'$, either because $s$
  is already some $(\psuccess,w)$, or because $s$ can reach
  $(q_0,\epsilon)$ via the cleaning gadget. As a consequence, all locations
  of $\L'$ are in $\Prom(\{\psuccess\})$, and then in
  $\Safe(\Prom(\{\psuccess\}))$.
  
  (vi) $\Longrightarrow$ (vii): trivial.
  
  (vii) $\Longrightarrow$ (i): obvious because $\Safe(A)\subseteq A$ for any
  set $A$ of locations.
\end{proof}

Using Lemma~\ref{lem-L'} and characterizations given by
Theorems~\ref{thm:eventually} and~\ref{thm:Buechi-decid} we have: 
\begin{align*}
\exists\U\ \Pr\nolimits_\U\bigl( (q_0,\epsilon)\models \Diamond
  \psuccess\bigr) = 1 & \text{ iff } q_0 \in \Prom(\{\psuccess\}) \tag{a.2}
\\
 & \text{ iff } \text{in $\L$, } q_0 \step{*} \paccept.
\\[1em]
\exists\U\ \Pr\nolimits_\U\bigl( (q_0,\epsilon)\models \Box\Diamond
  \psuccess\bigr) =1 & \text{ iff } q_0 \in \Safe(\Prom(\{\psuccess\})) \tag{b.2}
\\
& \text{ iff } \text{in $\L$, } q_0 \step{*} \paccept.
\\[1em]
\exists\U\ \Pr\nolimits_\U\bigl( (q_0,\epsilon)\models \Box\Diamond
  \neg \psuccess\bigr) =0 & \text{ iff } q_0 \in \Prom(\Safe(Q \setminus
  \{\psuccess\})) \tag{b.1}
\\
& \text{ iff } \text{in $\L$, } q_0 \step{*} \paccept.
\\[1em]
\exists\U\ \Pr\nolimits_\U\bigl( (q_0,\epsilon)\models \Box\Diamond
  \neg \psuccess\bigr) \ls 1 & \text{ iff } q_0 \step{*} Q \setminus \{\psuccess\} \tag{b.3}
\\
& \text{ iff } \text{in $\L$, } q_0 \step{*} \paccept.
\end{align*}

Thus, $q_0 \step{*} \paccept$, a non-primitive recursive problem,
reduces to instances of $(a.1)$, $(b.2)$, $(b.1)$ and $(b.3)$.
  
\begin{figure}[htbp]
\centering
\includegraphics{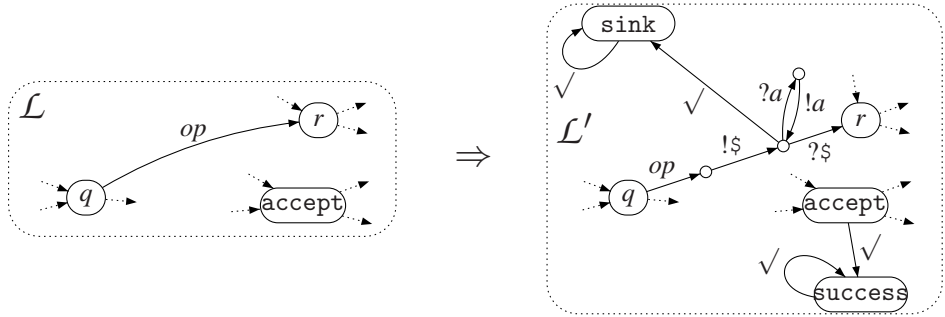}
\caption{Associating $\L'$ with an arbitrary LCS $\L$ for case $(a.3)$}
\label{fig-reduc-4}
\end{figure}

We now prove case (a.3) of Theorem~\ref{non-prim}, using the
reduction described in Fig.~\ref{fig-reduc-4}.

Here, with some LCS 
$\L$ as before,  we associate an LCS $\L'$ by adding two
special locations $\psink$ and $\psuccess$. As in the previous
reduction, $\psuccess$ is directly reachable from $\paccept$ by an
internal action $\surd$, and one can loop on $\psuccess$.

Now, each transition rule $\delta:q \step{\op} r$ in $\L$ is translated in
$\L'$ under the form
$q\step{\op}l_\delta\step{!\ttdol}l'_\delta\step{?\ttdol}r$, using two
intermediate locations $l_\delta$ and $l'_\delta$, and a new message
$\ttdol\not\in\Mess$.  Thus, moving from $q$ to $r$ in $\L'$ requires that
one removes the extra $\ttdol$ that has just been inserted. This is
obtained by a full rotation of the channel contents, using extra rules
$l'_\delta\step{?a}\_\step{!a}l'_\delta$ that exist for each $a\in\Mess$.
Finally, in case of deadlocks induced by message losses, one can go to the
$\psink$ location.

The purpose of this reduction is to ensure that $\paccept$ and
$\psuccess$ are the only locations from which one  can surely, i.e., with
probability one, reach $\psuccess$. For all other locations, the channel
may become empty along the way to $\paccept$, forcing the system to go to $\psink$.

\begin{lemma}
  In $\N=(\L',\tau)$ the following assertions are equivalent:
\begin{enumerate}
\item[(1)] $\exists \U\ \Pr_\U\bigl( (q_0,\epsilon)\models \Diamond
  \psink\bigr) \ls 1$,
\item[(2)] $q_0 \step{*}_{[\neg \psink]} \Safe(Q \setminus \{\psink\})$,
\item[(3)] $q_0 \step{*}_{[\neg \psink]} \{\paccept,\psuccess\}$,
\item[(4)] $q_0 \step{*}_{[\L]} \{\paccept\}$.
\end{enumerate}
where here again ``$(q_0,\epsilon) \step{*}_{[\L]}\cdots$'' means that
the path only visits original locations from $\L$.
\end{lemma}

\begin{proof}
  The equivalence between (1) and (2) is given by case (c) of
  Theorem~\ref{thm:eventually}. Then we show that $\Safe(Q \setminus
  \{\psink\}) = \{\paccept,\psuccess\}$. First $\{\paccept,\psuccess\}
  \subseteq \Safe(Q \setminus \{\psink\})$ because from $\paccept$ and
  $\psuccess$ one can loop forever in $\psuccess$ which is in $Q
  \setminus \{\psink\}$. On the other hand, if we consider another
  location $q$ different from $\psink$ (neither $\psuccess$ nor
  $\paccept$) because of the reading operation between $l'_\delta$ and
  $r$, there is a non-zero probability for the system to lose the
  message $\ttdol$ and be forced to go to $\psink$. Hence $\Safe(Q
  \setminus \{\psink\})$ is exactly $\{\paccept,\psuccess\}$.
  Equivalences of (2) with (3) and (4) follow from this equality.
\end{proof}

Thus the non-primitive recursive problem ``does $(q_0,\varepsilon)
\step{*} \paccept$'' reduces to a special instance of problem $(a.3)$
in Theorem~\ref{non-prim}.

\subsection{Undecidability} 
\label{ssec-undecidability}

\subsubsection{An undecidability result for repeated eventually properties} 
\label{sec:undec-1}

We will now combine the cleaning gadget with an arbitrary lossy
channel system to get a reduction from the boundedness problem for
LCS's to the question whether a single B\"uchi constraint $\Box
\Diamond A$ holds with positive probability under some scheduler.
Recall that an LCS $\L$ is \emph{bounded} (also space-bounded) for a
given a starting configuration if the set of reachable configurations
is finite.

\begin{theorem}[(Single B\"uchi property, positive probability)]
\label{thm:undec-single-Buechi-pos-prob}
The problem,\\  given $\N$ a NPLCS, $q$ a location, and $A$ a set of
locations, whether there exists a scheduler $\U$ such that
$\Pr_\U\bigl((q,\epsilon) \models \Box \Diamond A \bigr) \gr 0$, is
undecidable.
\end{theorem}

\begin{figure}[htbp]
\centering
\includegraphics{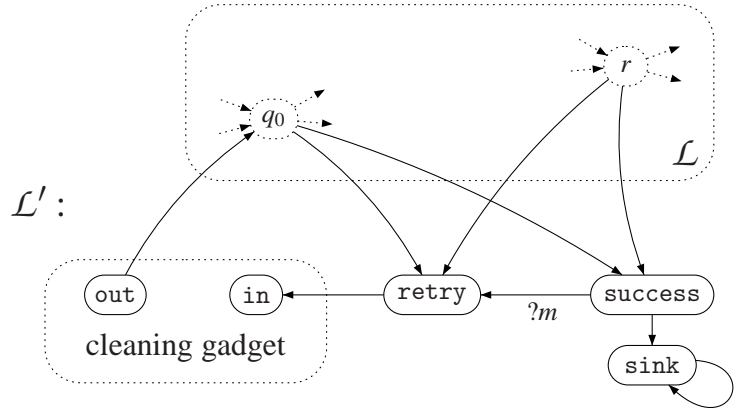}
\caption{The LCS $\L'$ associated with $\L$ in proof of Theorem~\ref{thm:undec-single-Buechi-pos-prob}}
\label{fig-undec-1}
\end{figure}

The remainder of this subsection is concerned with the proof of
Theorem~\ref{thm:undec-single-Buechi-pos-prob}.
Let $\L=(Q,\{c\},\Mess,\Delta)$ be a LCS with a single
channel $c$ and a designated initial configuration $(q,\epsilon)$. We
modify $\L$ by adding the cleaning gadget and two locations:
$\psuccess$ and $\psink$. We also add rules allowing to jump from
every ``original'' location in $Q$ to $\pretry$ or $\psuccess$.
When in
$\psuccess$, one can move to $\pretry$ with a read or move to $\psink$
which cannot be left.  When in $\pretry$, one can go back to
$(q,\epsilon)$ through the cleaning gadget.
The whole construction is depicted in Fig.~\ref{fig-undec-1}.

Let $\L'$ be the resulting LCS which we consider as an NPLCS with some
fault rate $\tau$: $\N=(\L',\tau)$. Since the cleaning gadget lets one
go back to the initial configuration of $\L$, any behavior of $\L'$ is
a succession of behaviors of $\L$ separated by visits to the
additional locations. The idea of this construction is the following:
if $\L$ is bounded, then even the best scheduler cannot visit
$\psuccess$ infinitely often without ending up in $\psink$ almost
surely. However, if the system $\L$ is bounded, some infinite memory
scheduler can achieve this. These ideas are formalized in
Propositions~\ref{prop-undec-1-rl} and \ref{prop-undec-1-lr}.

\begin{proposition}
\label{prop-undec-1-rl}
Assume that $\L$ starting from $(q,\varepsilon)$ is bounded. Then,
for all schedulers $\U$ for $\N = (\L',\tau)$,
$\Pr_\U\bigl((q,\epsilon) \models \Box\Diamond {\psuccess}\bigr) =0$.
\end{proposition}
\begin{proof}
  Let $\U$ be any scheduler for $\N$ and consider the $\U$-paths that
  visit $\psuccess$ infinitely often. Let $\pi$ be one such path:
  either $\pi$ jumps from $\L$ to $\psuccess$ infinitely many times,
  or it ends up in $\psink$. In the last case, $\pi$ does not satisfy
  $\Box\Diamond\psuccess$. In the first case, and since $\L$
  is bounded, $\pi$ can only jump to $\psuccess$ from finitely many
  different configurations.  Hence, for each such jump, the
  probability that it ends in $(\psuccess,\epsilon)$ is at least $\tau^m$,
  where $m$ is the size of the largest reachable configuration in
  $\L$.  Therefore, the configurations $(\psuccess,\epsilon)$ will be
  visited almost surely. As only the transition rule
  \[
                       \psuccess \step{\surd} \psink
  \]
  is enabled in $(\psuccess,\epsilon)$,
  with probability 1 the location $\psink$ is eventually reached.
  Since $\psuccess$ is not reachable from $\psink$, the property $\Box
  \Diamond \psuccess$ holds with zero probability.
\end{proof}

\begin{proposition}
\label{prop-undec-1-lr}
Assume that $\L$ starting from $(q,\varepsilon)$ is unbounded.
Then, there exists a scheduler $\U$ for $\N = (\L',\tau)$ with
$\Pr_\U\bigl( (q,\epsilon) \models \Box\Diamond {\psuccess} \bigr) \gr
0$.
\end{proposition}

\begin{proof}
  We describe the required scheduler $\U$.  Because $\L$ is
  unbounded, we can pick a sequence
  $\bigl((r_n,w_n)\bigr)_{n=1,2,\ldots}$ of reachable configurations
  such that $|w_n|\geq n$.  The scheduler works in phases numbered
  $1,2,\ldots$ When phase $n$ starts, $\U$ is in the initial
  configuration $(q,\epsilon)$ and tries to reach $(r_n,w_n)$. In
  principle, this can be achieved (since $(r_n,w_n)$ is reachable),
  but it requires that the right messages are lost at the right times.
  These losses are probabilistic and $\U$ cannot control them. Thus
  $\U$ aims for $(r_n,w_n)$ and hopes for the best.  It goes on
  according to plan as long as losses occur as hoped. When a ``wrong''
  loss occurs, $\U$ resigns temporarily, jumps directly to $\pretry$,
  reaches the initial configuration $(q,\epsilon)$ via the cleaning
  gadget, and then tries again to reach $(r_n,w_n)$.  When $(r_n,w_n)$
  is eventually reached (which will happen almost surely given enough
  retries), $\U$ jumps to $\psuccess$, from there to $\pretry$, and
  initiates phase $n+1$. With these successive phases, $\U$ tries to
  visit $\psuccess$ (and $\pretry$) an infinite number of times. We
  now show that it succeeds with nonzero probability.
  
  When moving from configuration $(r_n,w_n)$ to location $\psuccess$,
  there is a nonzero probability $\bfP_\lost(w_n,\epsilon)$ that all
  messages in the channel are lost, leaving us in $(\psuccess,\epsilon)$.
  When this happens, $\U$ is not able to initiate phase $n+1$ (moving
  from $\psuccess$ to $\pretry$ requires a nonempty channel). Instead
  $\U$ will move to $\psink$ and stay there forever. However, the
  probability for this exceptional behavior is strictly less than 1,
  as we have:
\[
  \Pr\nolimits_\U\bigl( (q,\epsilon) \models \Box\Diamond\psuccess
  \bigr) = \prod_{n=1}^{\infty} (1-\bfP_\lost(w_n,\epsilon)) \geq
  \prod_{n=1}^{\infty} (1-\tau^n) \gr 0.
\]
%
\end{proof}
Observe that the scheduler we constructed is recursive but not
finite-memory (since it records the index of the current phase).
\begin{remark}
  Proposition~\ref{prop-undec-1-lr} can be strengthened: if $\L$ is
  unbounded, then for all constant $c<1$, there exists a scheduler $\U$
  such that $\Pr_\U\bigl( (q,\epsilon) \models \Box\Diamond
  {\psuccess} \bigr) \gr c$. 
\end{remark}

\begin{corollary}
\label{cor-reduc-1}
Let $\L$ be a LCS. Then, $\L$ is unbounded if and only if there exists
a scheduler $\U$ for $\N=(\L',\tau)$ such that $\Pr_\U \bigl(
(q,\epsilon) \models \Box\Diamond {\psuccess} \bigr) \gr 0$.
\end{corollary}

This proves
Theorem~\ref{thm:undec-single-Buechi-pos-prob}
since it is undecidable whether a given LCS is bounded \cite{Mayr-unreliable}.

By duality we obtain the undecidability of the problem to check
whether $\Pr_\U\bigl((q,\epsilon) \models \Diamond\Box A\bigr)=1$ for all
schedulers $\U$ for a given NPLCS $\N$.

\subsubsection{Other undecidability results}
We now discuss the decidability of the problem which asks for a
scheduler $\U$ where $\Pr_\U\bigl((q,\epsilon) \models \varphi\bigr)$ is 1,
$\ls 1$, $=0$ or $\gr 0$ and where $\varphi$ is an LTL-formula.  We
begin with the special case of a strong fairness (Streett condition)
$\varphi = \bigwedge_{1 \leq i \leq n} (\Box \Diamond A_i \impl \Box
\Diamond B_i)$.  We will see that all variants of the qualitative
model checking problem for such Streett conditions are undecidable
when ranging over the full class of schedulers.  In particular, this
yields the undecidability of the LTL model checking problem when
considering all schedulers.  However, when we shrink our attention to
finite-memory schedulers qualitative model checking is decidable for
properties specified by Streett conditions or even $\omega$-regular formulas.

We first establish the undecidability results when ranging over all
schedulers. In fact, already a special kind of Streett properties
with the probabilistic satisfaction criterion ``almost surely'' cannot
be treated algorithmically:

\begin{lemma}
\label{lem-undec-Streett=1}
The problem, given NPLCS $\N$, sets of locations $A, B \subseteq Q$, and
location $q \in Q$, whether there exists a scheduler $\U$ with
\[
    \Pr\nolimits_\U \bigl((q,\epsilon) \models \Box \Diamond B \wedge
    \Diamond \Box A)\bigr) =1,
\]
is undecidable.
\end{lemma}

\begin{proof}
  The proof is again by a reduction from the boundedness problem for
  LCS as in section~\ref{sec:undec-1}.  Let $\L$ be an LCS. 
We build a new  LCS $\L'$  by combining $\L$
  with the cleaning gadget as shown in Fig.~\ref{fig-undec-2} (this is a
  variant of the previous construction). Let $\N=(\L',\tau)$.
\begin{figure}[htbp]
\centering
\includegraphics{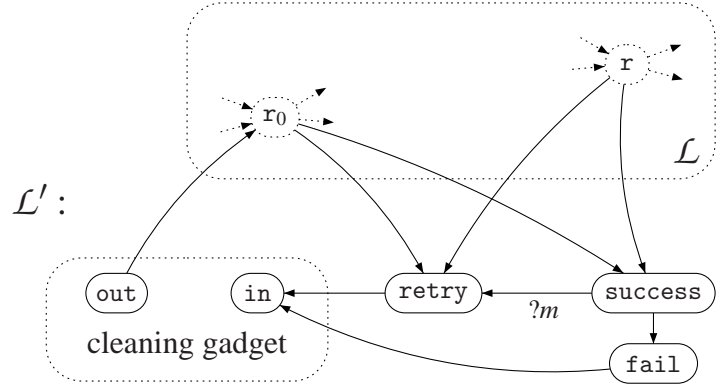}
\caption{The LCS $\L'$ associated with $\L$ in proof of Lemma~\ref{lem-undec-Streett=1}}
\label{fig-undec-2}
\end{figure}
There exists a scheduler $\U$ for $\N$ with
$\Pr_\U\bigl((q_0,\epsilon) \sat \Box\Diamond\psuccess \wedge
\Diamond\Box\neg\pfail \bigr)=1$ iff $\L$ is unbounded (starting from
$(q_0,\epsilon)$).

For these two constructions, the ``same'' scheduler is used in the
positive cases. For the second construction, the proof for the
positive case observes that
\[
  \Pr\nolimits_\U\bigl((q_0,\epsilon) \sat \Box \Diamond \psuccess \wedge
  \Diamond\Box \neg\pfail\bigr)= \lim_{n\rightarrow \infty}
  \prod_{k=n}^\infty(1-\tau^k)=1.
\]
where $n$ stands for the phase number from which $\pfail$ will not be
visited again.
\end{proof}

\begin{theorem}[(Streett properties)]
\label{thm:Streett-undec}
For the qualitative properties (a), \ldots, (d) below, the problem, given a NPLCS
$\N$, location $q \in Q$, and $2n$ sets of locations
$A_1,B_1,\ldots,$ $A_n,B_n\subseteq Q$, whether there exists a scheduler $\U$
such that
\begin{enumerate}
\item [(a)] $\Pr_\U\bigl( (q,\epsilon) \models
  \bigwedge\limits_{i=1}^n (\Box \Diamond A_i \impl \Box \Diamond
  B_i)\bigr) \gr 0$,
\item [(b)] $\Pr_\U\bigl( (q,\epsilon) \models
  \bigwedge\limits_{i=1}^n (\Box \Diamond A_i \impl \Box \Diamond
  B_i)\bigr) \ls 1$,
\item [(c)] $\Pr_\U\bigl( (q,\epsilon) \models
  \bigwedge\limits_{i=1}^n (\Box \Diamond A_i \impl \Box \Diamond
  B_i)\bigr) =1$,
\item [(d)] $\Pr_\U\bigl( (q,\epsilon) \models
  \bigwedge\limits_{i=1}^n (\Box \Diamond A_i \impl \Box \Diamond
  B_i)\bigr) = 0$,
\end{enumerate}
is undecidable.
\end{theorem}

\begin{proof} \ \
\begin{enumerate}
\item [(a)] follows immediately from
  Theorem~\ref{thm:undec-single-Buechi-pos-prob} as 
  $\bigwedge_{i=1}^n (\Box \Diamond A_i \impl \Box \Diamond B_i)$ agrees
  with $\Box \Diamond B$ if we take $n=1$, $A_1 = Q$ and $B_1=B$.

\item [(b)] We show that already the question whether there is some
  scheduler $\U$ with $\Pr_\U\bigl((q,\epsilon) \models \Box
  \Diamond A \impl \Box \Diamond B \bigr) \ls 1$ is undecidable where
  $A$ and $B$ are sets of locations.  This follows from
  Theorem~\ref{thm:undec-single-Buechi-pos-prob} and the fact 
  that for $B = \emptyset$
\begin{center}
\begin{tabular}{rcl}
    &   & $\Pr_\U\bigl((q,\epsilon) \models 
           \Box \Diamond A  \impl \Box \Diamond B\bigr) \ls 1$ 
\\[.3em]
iff &   & $\Pr_\U\bigl((q,\epsilon) \models 
           \neg (\Box \Diamond A  \impl \Box \Diamond B) \bigr) \gr 0$ 
\\[.3em]
iff &   & $\Pr_\U\bigl((q,\epsilon) \models 
            \Box \Diamond A  \wedge \underbrace{\Diamond \Box (Q \setminus B)}_{
              \equiv \true~\text{since $B = \emptyset$}}\bigr) \gr 0$ 
\\[1.9em]
iff &   & $\Pr_\U\bigl((q,\epsilon) \models 
           \Box \Diamond A \bigr) \gr 0$.
\end{tabular}
\end{center}

\item [(c)] follows by Lemma~\ref{lem-undec-Streett=1} with
  $n=2$, $A_1=Q$, $B_1=B$, $A_2= Q \setminus A$ and $B_2 = \emptyset$
  which yields
\begin{align*}
\bigwedge\limits_{1 \leq i \leq n}
 \bigl(\Box \Diamond A_i  \impl \Box \Diamond B_i\bigr)
& \;\;\equiv\;\;
\bigl(\underbrace{\Box \Diamond Q}_{\equiv \true}  \impl \Box \Diamond B\bigr)
\:\wedge\:
 \bigl(\Box \Diamond (Q\setminus A)  \impl 
    \underbrace{\Box \Diamond \emptyset}_{\equiv \false}\bigr)
\\
& \;\;\equiv\;\; \Box\Diamond B \:\wedge\: \Diamond \Box A.
\end{align*}

\item [(d)] We show the undecidability of the question whether
  $\Pr_\U\bigl((q,\epsilon) \models \Box \Diamond A \impl \Box
  \Diamond B\bigr) = 0$ for some $\U$ where $A,B \subseteq Q$ are
  given sets of locations.  This follows from  Lemma~\ref{lem-undec-Streett=1} and the fact that
\begin{center}
\begin{tabular}{rcl}
    &   & $\Pr_\U\bigl((q,\epsilon) \models \Box \Diamond A  \impl \Box \Diamond B \bigr) = 0$
\\
iff &   & $\Pr_\U\bigl((q,\epsilon) \models 
           \neg (\Box \Diamond A  \impl \Box \Diamond B) \bigr) = 1$
\\
iff &   & $\Pr_\U\bigl((q,\epsilon) \models 
           \Box \Diamond A  \wedge \Diamond \Box (Q \setminus B)\bigr) = 1$.
\end{tabular}
\end{center}
\end{enumerate}
\end{proof}

\begin{figure}[htbp]
\centering
\includegraphics{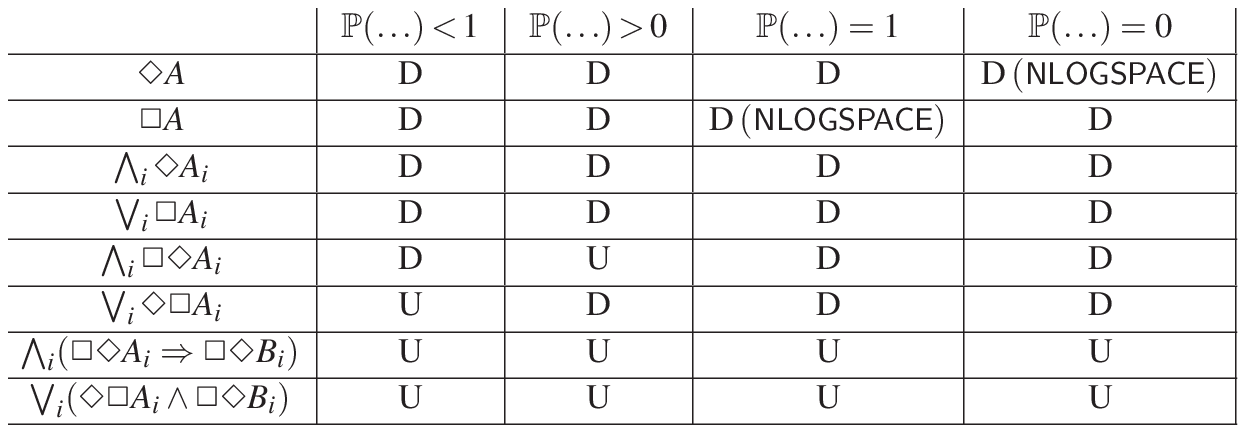}
\caption{(Un)Decidability of qualitative verification}
\label{fig-sum-table}
\end{figure}

Figure~\ref{fig-sum-table} summarizes the decidability and
undecidability results obtained so far.

\section{Restriction to finite-memory schedulers}
\label{sec-finite-mem}

In all decidable cases of section~\ref{sec-decid}, finite-memory
schedulers are sufficient. In this section we consider the problems of
section~\ref{sec-undec}, considering only finite-memory schedulers.
With this restriction, all problems are decidable.

We first give an immediate property of finite-memory schedulers which
will be used in the whole section.

\begin{proposition}
\label{property-fm-schedulers}
For any finite-memory scheduler $\U$ and any location $q$ we have:
\begin{enumerate}
\item [(a)] If $p$ is a location, $u$ a mode in $\U$ and if $T$
  denotes the set of all configurations $t$ that are reachable from
  $(p,\epsilon)_u$ by $\U$ then
\begin{gather*}
\Pr\nolimits_\U\bigl( (q,\epsilon) \models \Box \Diamond
(p,\epsilon)_u \bigr) = \Pr\nolimits_\U\bigl((q,\epsilon)
\models \bigwedge_{s \in T} \Box \Diamond s \bigr)
\end{gather*}
\item [(b)] $\Pr\nolimits_\U\bigl((q,\epsilon) \models \Box
  \Diamond A\bigr)= \Pr\nolimits_\U\bigl((q,\epsilon) \models \Box
  \Diamond A_\epsilon \bigr)$.
\end{enumerate}
\end{proposition}

\begin{proof}
  (a) If configuration $s_u$ in the Markov chain $\MC_\U$ is visited
  infinitely often then almost surely all direct successors of $s_u$
  are visited infinitely often too.  We now may repeat this argument
  for the direct successors of the direct successors of $s_u$, and so
  on.  We obtain that almost surely all configurations that are
  reachable from $s_u$ are visited infinitely often, provided that
  $s_u$ is visited infinitely often.
  
  (b) follows from (a) using the fact that the set of all $(p,\epsilon)_u$
  for $p$ a location and $u$ a mode of $\U$, is a finite attractor, and
  observing that if $(a,w)$ is reachable within one step from configuration
  $s$ then so is $(a,\epsilon)$ as all messages can be lost.
\end{proof}

Observe that the existence of a scheduler $\U$ for which a Büchi
property holds with positive probability, does not imply the existence
of a finite-memory scheduler with the same property. This is a
consequence of
Theorem~\ref{thm:undec-single-Buechi-pos-prob} and the
next Theorem (\ref{thm:Buechi-pos-prob}).

\begin{theorem}[(Generalized B\"uchi, positive probability)]
\label{thm:Buechi-pos-prob}
The problem, given NPLCS $\N$, location $q \in Q$, and sets of locations
$A_1,\ldots,A_{n}\subseteq Q$, whether there exists a \emph{finite-memory
scheduler} $\U$ such that $\Pr_\U\bigl((q,\epsilon) \models \bigwedge_{1
\leq i \leq n} \Box \Diamond A_i \bigr) \gr 0$, is decidable.
\end{theorem}

\begin{proof}
We show that the following statements (1) and (2) are equivalent:
\begin{enumerate}
\item [(1)]
there exists a finite-memory scheduler
$\U$ such that
$\Pr_\U\bigl((q,\epsilon) \models 
  \bigwedge\limits_{1 \leq i \leq n} \Box \Diamond A_i \bigr) \gr 0$.
\item [(2)] there exists a location $x \in Q$ such that
\begin{enumerate}
\item [(2.1)] $(q,\epsilon) \step{*} (x,\epsilon)$
\item [(2.2)] 
    there is a finite-memory scheduler $\V$ with
    $\Pr_{\V}\bigl((x,\epsilon) \models \bigwedge\limits_{1 \leq i \leq n}
  \Box \Diamond A_i\bigr)=1$
\end{enumerate}
\end{enumerate}

This will prove Theorem~\ref{thm:Buechi-pos-prob} since by
Theorem~\ref{thm:Buechi-decid} (a), there is an algorithmic way to
compute the set $X$ of locations $x$ such that
$\Pr_{\V}\bigl((x,\epsilon) \models \bigwedge_{1 \leq i \leq n} \Box
\Diamond A_i\bigr)=1$ for some (finite-memory) scheduler $\V$.  We
then may check (2.1) by an ordinary reachability analysis in the
underlying LCS.

Let us show the equivalence of (1) and (2).

(1) $\Longrightarrow$ (2):
Let $\U$ be a finite-memory scheduler as in (1).
The finite-attractor property and
Proposition~\ref{property-fm-schedulers} yield that there is some 
location $x$ and mode $u$ of $\U$ 
with
\begin{gather*}
  \Pr\nolimits_\U\bigl( (q,\epsilon) \models \bigwedge_{1\leq i \leq
    n} \Box \Diamond A_i \wedge \bigwedge_{t \in T} \Box \Diamond t
  \bigr) \gr 0
\end{gather*}
where $T$ is the set of configurations that are reachable from
$(x,\epsilon)_u$ under $\U$. Using definition of $T$, this yields $T
\cap A_i \neq \emptyset$ for $1 \leq i \leq n$.  Thus, scheduler $\U$
starting in $(x,\epsilon)$ in mode $u$ visits almost surely any
configuration in $T$ infinitely often. Hence, it visits any set $A_i$
for $i=1,\ldots,n$, infinitely often (with probability one). That is:
\begin{gather*}
  \Pr\nolimits_\U\bigl( (x,\epsilon)_u \models \bigwedge_{1\leq i \leq
    n} \Box \Diamond A_i \bigr) =1,
\end{gather*}
and (2) holds.

(2) $\Longrightarrow$ (1): Let $q$, $x$ and $\V$ be as in (2).  We
define $\U$ as the finite-memory scheduler that generates with
positive probability a path from $(q,\epsilon)$ to $(x,\epsilon)$ and
behaves as $\V$ from $(x,\epsilon)$ on.  Clearly, we then have
$\Pr_\U\bigl( (q,\epsilon) \models \bigwedge_{1\leq i \leq n} \Box
\Diamond A_i\bigr) \gr 0$.
\end{proof}

We now present algorithms for the four variants of qualitative model
checking of Streett properties for NPLCS's when ranging over
finite-memory schedulers.

\begin{theorem}[(Streett properties)]
\label{thm:streett-fm}
For qualitative properties (a), \ldots, (d), the problem, given NPLCS $\N$,
location $q \in Q$, and $2n$ sets of locations
$A_1,B_1,\ldots,A_n,B_n\subseteq Q$, whether there exists a
\emph{finite-memory} scheduler $\U$ satisfying
\begin{itemize}
\item [\emph{(a)}] $\Pr_\U\bigl((q,\epsilon) \models \bigwedge\limits_{1
    \leq i \leq n} (\Box \Diamond A_i \impl \Box \Diamond B_i)\bigr) \ls 1$,
\item [\emph{(b)}] $\Pr_\U\bigl((q,\epsilon) \models \bigwedge\limits_{1
    \leq i \leq n} (\Box \Diamond A_i \impl \Box \Diamond B_i)\bigr) \gr 0$,
\item [\emph{(c)}] $\Pr_\U\bigl((q,\epsilon) \models \bigwedge\limits_{1
    \leq i \leq n} (\Box \Diamond A_i \impl \Box \Diamond B_i)\bigr) = 1$, 
\item [\emph{(d})] $\Pr_\U\bigl((q,\epsilon) \models \bigwedge\limits_{1
    \leq i \leq n} (\Box \Diamond A_i \impl \Box \Diamond B_i)\bigr) = 0$,
\end{itemize}
is decidable.
\end{theorem}

We prove each assertion in the rest of this section.

\textbf{ad (a)} of Theorem~\ref{thm:streett-fm}: $\Pr_\U\bigl((q,\epsilon)
\models \bigwedge\limits_{1 \leq i \leq n} (\Box \Diamond A_i \impl \Box
\Diamond B_i)\bigr) \ls 1$.

Let us consider the dual problem whether, for all finite-memory
schedulers $\U$,
\[
\Pr\nolimits_\U\bigl( (q,\epsilon) \models \bigwedge\limits_{i=1}^n
(\Box \Diamond A_i \impl \Box \Diamond B_i)\bigr) = 1
\]
Clearly, the above holds iff
\[
\Pr\nolimits_\U\bigl( (q,\epsilon) \models \Box \Diamond A_i \impl
\Box \Diamond B_i \bigr) = 1
\]
for all finite-memory schedulers $\U$ and all indices $i =1,\ldots,n$.
Thus, it suffices to present an algorithm that solves the problem whether
$\Pr_\U\bigl((q,\epsilon) \models \Box \Diamond A \impl \Box \Diamond B
\bigr) = 1$ for all finite-memory schedulers $\U$ where $A$ and $B$ are
given sets of locations. The latter is equivalent to the non-existence of a
finite-memory scheduler $\U$ such that 
\[
\Pr\nolimits_\U\bigl((q,\epsilon) \models \Box \Diamond A \wedge
\Diamond \Box (Q \setminus B)\bigr) \gr 0.
\]
We now explain how to check this condition algorithmically. Let $\N'$
be the NPLCS that arises from $\N$ by removing all locations $b \in
B$.  To ensure that any configuration has at least one outgoing
transition, we add a new location $\pfail$ with
\begin{itemize}
\item
a self-loop
$\pfail \step{\surd} \pfail$ and
\item transition rules $p \step{\op} \pfail$ if $p \step{\op}
  b$ for some location $b \in B$.
\end{itemize}
Using Theorem~\ref{thm:Buechi-pos-prob}, we can compute the set $P$ of
locations $p\in Q\setminus B$ such that there is a finite-memory
scheduler $\U'$ for $\N'$ with $\Pr_{\U'}\bigl((p,\epsilon) \models
\Box \Diamond A \bigr) \gr 0$.  That is,
\begin{gather*}
P = \left\{
p\in Q \left| \begin{array}{c}
         \text{there is some finite-memory scheduler $\U$ for $\N$} \\
         \text{with }\Pr_{\U}\bigl((p,\epsilon)\models\Box\Diamond A \wedge\Box\neg B \bigr) \gr 0 
\end{array}
\right. 
\right\}.
\end{gather*}
We show the equivalence of the following two statements:
\begin{description}
\item [(1)] $\Pr_\U\bigl((q,\epsilon) \models \Box \Diamond A
  \wedge \Diamond \Box \neg B \bigr) \gr 0$ for some finite-memory
  scheduler $\U$ for $\N$,
\item [(2)] $\Pr_\V\bigl((q,\epsilon) \models \Diamond P\bigr) \gr
  0$ for some finite-memory scheduler $\V$ for $\N$.
\end{description}
(1) $\Longrightarrow$ (2): Let $\U$ be a finite-memory scheduler as in
(1).  By Proposition~\ref{property-fm-schedulers}, we may conclude
that there exists a location $a \in A$ and a mode $u$ such that
\[
\Pr\nolimits_\U\bigl((q,\epsilon) \models \bigwedge_{s \in T} \Box
\Diamond s \wedge \Diamond \Box \neg B \bigr) \gr 0
\]
where $T$ is the set of states that are reachable in the Markov chain
$\MC_\U$ from $(a,\epsilon)_u$, i.e., from configuration
$(a,\epsilon)$ in mode $u$.  We then have $T \cap B = \emptyset$ and
\[
\Pr\nolimits_\U\bigl((a,\epsilon)_u \models \bigwedge_{s \in T} \Box
\Diamond s \wedge \Diamond \Box \neg B \bigr) =
\Pr\nolimits_\U\bigl((a,\epsilon)_u \models \Box \Diamond A \wedge
\Box \neg B \bigr) =1.
\]
Hence, $a \in P$ and $\Pr_\U\bigl((q,\epsilon) \models \Diamond
P\bigr) \gr 0$.

(2) $\Longrightarrow$ (1): Let $\V$ be a finite-memory scheduler as in
(2).  For any location $p \in P$, there is a finite-memory
scheduler $\U_p$ such that
\[
\Pr\nolimits_{\U_p}\bigl((p,\epsilon) \models \Box \Diamond A
\wedge \Box \neg B \bigr) \gr 0.
\]
 We now may compose $\V$ and the schedulers $\U_p$ to obtain a
 finite-memory scheduler $\U$ which first mimics $\V$ until we reach a
 configuration $(p,\epsilon)$ for some $p \in P$ (which happens
 with positive probability) and which then behaves as $\U_p$.
 Clearly, we then have $\Pr_\U\bigl((q,\epsilon) \models \Box
 \Diamond A \wedge \Diamond \Box \neg B \bigr) \gr 0$.


\textbf{ad (b)} of Theorem~\ref{thm:streett-fm}: $\Pr_\U\bigl((q,\epsilon)
\models \bigwedge\limits_{1 \leq i \leq n} (\Box \Diamond A_i \impl \Box
\Diamond B_i)\bigr) \gr 0$.
 
Let $I \subseteq \{1,\ldots,n\}$ and $\N_{~I}$ be the NPLCS that arises
from $\N$ by removing the locations $b \in A_i$ where $i \in
\{1,\ldots,n\} \setminus I$, and adding a new location $\pfail$ as in
the proof of ad (a) (of the present Theorem).

Let $C_I$ be the set of locations $z \in Q$ such that $\Pr_\U \bigl(
(z,\epsilon) \models \bigwedge_{i \in I} \Box \Diamond B_i\bigr) =1$
for some (finite-memory) scheduler $\U$ for $\N_{~I}$.  Note that under
such a scheduler $\U$ the new location $\pfail$ is not reachable from
$(c,\epsilon)$.  Then, we have $z \in C_I$ iff there exists a
finite-memory scheduler $\U_z$ for the original NPLCS $\N$ with
\[
\Pr\nolimits_{\U_z} \bigl( (z,\epsilon) \models \bigwedge\limits_{i
  \in I} \Box \Diamond B_i \ \wedge \!\!\bigwedge\limits_{\stackrel{i
    \in \{1,\ldots,n\}}{ i \notin I}} \!\!\!\Box \neg A_i \bigr) =1.
\]
In particular, $\Pr_{\U_z}\bigl((z,\epsilon) \models
\!\!\bigwedge\limits_{i=1}^n (\Box \Diamond A_i \impl \Box \Diamond
B_i)\bigr) =1$ for all $z \in Z$.

The $C_I$'s can be computed with the technique explained in the proof
of Theorem~\ref{thm:Buechi-decid} (part (a)). Let $C$ be the union of
all $C_I$'s.  Then, the following statements are equivalent:
\begin{description}
\item [(1)] $C$ is reachable from $(q,\epsilon)$
\item [(2)] $\Pr_\U\bigl((q,\epsilon) \models
  \bigwedge\limits_{i=1}^n (\Box \Diamond A_i \impl \Box \Diamond
  B_i)\bigr) \gr 0$ for some finite-memory scheduler $\U$.
\end{description}
(1) $\Longrightarrow$ (2): Let us assume that $C$ is reachable from
$(q,\epsilon)$.  Then, there is a memoryless scheduler $\U_{\it
  init}$ such that $\Pr_{\U_{\it init}}\bigl( (q,\epsilon) \models
\Diamond C\bigr) \gr 0$.  Hence, there is some $z \in C$ such that
\[
\Pr\nolimits_{\U_{\it init}}\bigl( (q,\epsilon) \models \Diamond
(z,\epsilon)\bigr) \gr 0.
\]
 We then may combine $\U_{{\it init}}$ and $\U_z$ to obtain a
 finite-memory scheduler $\U$ with the desired property.

(2) $\Longrightarrow$ (1): Let us now assume that $\U$ is a
finite-memory scheduler such that
\[
\Pr\nolimits_\U\bigl((q,\epsilon) \models
\!\!\bigwedge\limits_{i=1}^n (\Box \Diamond A_i \impl \Box \Diamond
B_i)\bigr) \gr 0.
\]
Then, there is some $I \subseteq \{1,\ldots,n\}$ such that
\[
\Pr\nolimits_\U\bigl((q,\epsilon) \models \bigwedge\limits_{i \in
  I} \Box \Diamond B_i \wedge \bigwedge\limits_{i \notin I} \Diamond
\Box \neg A_i\bigr) \gr 0.
\]
The finite-attractor property yields the existence of some location
$z$ and a mode $u$ of $\U$ such that
\[
\Pr\nolimits_\U\bigl((q,\epsilon) \models \Box \Diamond
(z,\epsilon)_u \wedge \bigwedge\limits_{i \in I} \Box \Diamond B_i
\wedge \bigwedge\limits_{i \notin I} \Diamond \Box \neg A_i \bigr) \gr
0.
\]
As visiting $(z,\epsilon)_u$ infinitely often ensures that almost
surely all configurations that are reachable from $(z,\epsilon)_u$
are visited infinitely often too (see
Proposition~\ref{property-fm-schedulers}), we obtain 
\[
\Pr\nolimits_\U \bigl((z,\epsilon)_u \models \bigwedge\limits_{i
  \in I} \Box \Diamond B_i \wedge \bigwedge\limits_{i \notin I} \Box
\neg A_i \bigr) =1.
\]
Hence, $z \in C_I \subseteq C$. This yields that $C$ is reachable from
$(q,\epsilon)$.


\textbf{ad (c)} of Theorem~\ref{thm:streett-fm}: $\Pr_\U\bigl((q,\epsilon)
\models \bigwedge\limits_{1 \leq i \leq n} (\Box \Diamond A_i \impl \Box
\Diamond B_i)\bigr) = 1$.

Let $C$ be as in the proof of ad (b). We establish the equivalence of
the following statements:
\begin{description}
\item [(1)] $\Pr_\U \bigl((q,\epsilon) \models
  \bigwedge\limits_{i=1}^n (\Box \Diamond A_i \impl \Box \Diamond
  B_i)\bigr) =1$ for some finite-memory scheduler $\U$,
\item [(2)] $\Pr_\V\bigl((q,\epsilon) \models \Diamond C\bigr)=1$
  for some finite-memory scheduler $\V$.
\end{description}

(2) $\Longrightarrow$ (1): Let $\V$ be a finite-memory scheduler such
that $\Pr_\V\bigl((q,\epsilon) \models \Diamond C\bigr)=1$.  For $z
\in C$, let $\U_z$ be a finite-memory scheduler as in the proof of
assertion (b).  That is such that
\[
\Pr\nolimits_{\U_z}\bigl((z,\epsilon) \models
\bigwedge\limits_{i=1}^n (\Box \Diamond A_i \impl \Box \Diamond
B_i)\bigr) =1.
\]
Then, we may compose $\V$ and the finite-memory schedulers $\U_z$ to
obtain a finite-memory scheduler $\U$ such that
\[
\Pr\nolimits_\U\bigl((q,\epsilon) \models \bigwedge\limits_{i=1}^n
(\Box \Diamond A_i \impl \Box \Diamond B_i) \bigr) =1.
\]
Starting in $(q,\epsilon)$, $\U$ mimics $\V$ until a configuration
$(z,w)$ with $z \in C$ is reached (this happens with probability 1).
Then, for $w \neq \epsilon$, $\U$ chooses the transition rule
\[
                         \delta_z = z \step{\op}y
\]
that $\U_z$ chooses for $(z,\epsilon)$ in its initial mode.  Note that
$\delta_z$ is enabled in $(z,w)$, and all successors of $(z,w)$ under
$\delta_z$ have the form $(y,w')$ for some channel valuation $w'$.
Moreover, location $y$ belongs to $C$ as $\U_z$ induces a scheduler
$\U'_z$ with
\[
\Pr\nolimits_{\U'_z}\bigl((y,\epsilon) \models
\bigwedge\limits_{i=1}^n (\Box \Diamond A_i \impl \Box \Diamond
B_i)\bigr) =1.
\]
Hence, if $w' \neq \varepsilon$ then $\U$ may choose the transition
rule $\delta_y$ that $\U_y$ chooses for its starting configuration
$(y,\epsilon)$.  $\U$ continues in that way until it reaches a
configuration $(x,\epsilon)$.  (The finite-attractor property ensures
that this happens with probability 1.)  The above construction ensures
that $x \in C$.  After reaching $(x,\epsilon)$, $\U$ behaves as
$\U_x$, ensuring that $\bigwedge\limits_{i=1}^{n} (\Box \Diamond A_i
\impl \Box \Diamond B_i)$ holds almost surely.

(1) $\Longrightarrow$ (2): Let $\U$ be a finite-memory scheduler such
that
\[
\Pr\nolimits_\U\bigl((q,\epsilon) \models \bigwedge\limits_{i=1}^n
(\Box \Diamond A_i \impl \Box \Diamond B_i)\bigr) =1.
\]
We show that:
\begin{equation}
\label{eq-star}
\tag{*}
\text{For any location $p \in Q$: if
  $\Pr\nolimits_\U\bigl((q,\epsilon) \models \Box \Diamond
  (p,\epsilon)\bigr)\gr 0$ then $p \in C$.}
\end{equation}
Using the fact that $\Pr_\U\bigl( (q,\epsilon) \models \bigvee_{p \in Q}
\Box \Diamond (p,\epsilon)\bigr) =1$, \eqref{eq-star} yields
$\Pr_\U\bigl((q,\epsilon) \models \Diamond C\bigr)=1$.

\begin{proof}[(of~\eqref{eq-star})]
  Assume that $u$ is a mode in $\U$ such that $\Pr_\U\bigl(
  (q,\epsilon) \models \Box \Diamond (p,\epsilon)_u\bigr) \gr 0$.  Let
  $T$ be the set of states that are reachable from $(p,\epsilon)_u$ in
  the Markov chain for $\U$.  Then, by
  Proposition~\ref{property-fm-schedulers}:
\[
\Pr\nolimits_\U\bigl((q,\epsilon) \models \bigwedge\limits_{t \in
  T} \Box \Diamond t \bigr)\gr 0.
\]
Hence,
\[
\Pr\nolimits_\U\bigl((q,\epsilon) \models \bigwedge_{t \in T} \Box
\Diamond t \wedge \bigwedge_{i=1}^n (\Box \Diamond A_i \impl \Box
\Diamond B_i)\bigr)\gr 0.
\]
Let $I$ be the set of indices $i \in \{1,\ldots,n\}$ such that $T \cap A_i
\neq \emptyset$. Then, $T \cap B_i \neq \emptyset$ for all $i \in I$.
Hence, 
\[
\Pr\nolimits_\U\bigl((p,\epsilon)_u \models \bigwedge_{i \in I} \Box
\Diamond B_i \wedge \bigwedge_{i \notin I} \Box \neg A_i\bigr) = 1.
\]
Thus, $p \in C_I \subseteq C$.
\end{proof}


\textbf{ad (d)} of Theorem~\ref{thm:streett-fm}: $\Pr_\U\bigl((q,\epsilon)
\models \bigwedge\limits_{1 \leq i \leq n} (\Box \Diamond A_i \impl \Box
\Diamond B_i)\bigr) =0$.

We deal with the negation of the Streett formula:
\[
\Pr\nolimits_\U\bigl((q,\epsilon) \models \!\bigwedge_{i=1}^{n}
(\Box \Diamond A_i \impl \Box \Diamond B_i)\bigr) = 0 \;\;\text{iff}\;\;
\Pr\nolimits_\U\bigl((q,\epsilon) \models \!\bigvee_{i=1}^{n} (\Box
\Diamond A_i \wedge \Diamond \Box \neg B_i)\bigr) = 1.
\]
Thus, it suffices to establish the decidability of the question
whether there is a finite-memory scheduler $\U$ with
$\Pr_\U\bigl((q,\epsilon) \models \bigvee\limits_{i=1}^n (\Box
\Diamond A_i \wedge \Diamond \Box \neg B_i)\bigr) = 1$.

For $i \in \{1,\ldots,n\}$, let $\N_{~i}$ be the NPLCS that arises
from $\N$ by removing all locations in $B_i$, possibly adding a new
location $\pfail$ (as in the proof of case (a)).  Let $C_i$ be the set
of locations $z \in Q$ such that there exists a scheduler $\U_i$ for
$\N_{~i}$ with
\[
\Pr\nolimits_{\U_i}\bigl((z,\epsilon) \models \Box \Diamond
A_i\bigr) =1.
\]
The set $C_i$ can be computed with the techniques sketched in
Theorem~\ref{thm:Buechi-decid} ad (a).  Then, $z \in C_i$ iff there
exists a scheduler $\U_i$ for the original NPLCS $\N$ with
\[
  \Pr\nolimits_{\U_i}\bigl((z,\epsilon) \models \Box \Diamond A_i
  \wedge \Box \neg B_i\bigr) =1.
\]
  Let $C = C_1 \cup \cdots \cup C_n$.  Then, the following two
  statements are equivalent:
\begin{description}
\item [(1)] There is a finite-memory scheduler $\U$ with
  $\Pr_\U\bigl((q,\epsilon) \models \bigvee\limits_{1 \leq i \leq
    n} \!\!  (\Box \Diamond A_i \wedge \Diamond \Box \neg B_i)\bigr) =
  1$.
\item [(2)] There is a scheduler $\V$ with
  $\Pr_\V\bigl((q,\epsilon) \models \Diamond C\bigr)=1$.
\end{description}
(1) $\Longrightarrow$ (2): Let $\U$ be as in (1).  Assume
$\Pr_\U\bigl((q,\epsilon) \models \Diamond C\bigr)\ls 1$.  Then,
\[
\Pr\nolimits_\U\bigl( (q,\epsilon) \models \Box (Q \setminus C)
\bigr) \gr 0.
\]
By the finite attractor property there exists a location $x$ such that
\[
\Pr\nolimits_\U\bigl( (q,\epsilon) \models \Box \Diamond
  (x,\epsilon) \wedge \Box (Q \setminus C) \bigr) \gr 0.
\]
  As $\U$ is finite-memory there is a mode $u$ of $\U$ such that
  the above condition holds for $(x,\epsilon)$ in mode $u$, that
  is,
\[
\Pr\nolimits_\U\bigl( (q,\epsilon) \models \Box \Diamond (x,\epsilon)_u
  \wedge \Box (Q \setminus C) \bigr)
  \gr 0.
\]
Let $T$ be the set of configurations that are reachable from
$(x,\epsilon)_u$ in the Markov chain induced by $\U$, $\MC_\U$.  Then,
almost surely $\U$ visits all configurations in $T$ infinitely often
when starting in $(x,\epsilon)$ in mode $u$.  We then have $T \cap \{
(z,w) \in \Conf \mid z \in C\} = \emptyset$, and hence,
\[
  T \cap \bigl\{ (z,w) \in \Conf : z \in C_i\bigr\} = \emptyset, \ \ 
  i=1,\ldots,n,
\]
  which gives us $T \cap A_i = \emptyset$ or $T \cap B_i \neq
  \emptyset$ for any $i \in \{1,\ldots,n\}$.  But then,
\[
\Pr\nolimits_\U\bigl((x,\epsilon)_u \models \bigvee\limits_{i=1}^n
(\Box \Diamond A_i \wedge \Diamond \Box B_i)\bigr) = 0.
\]
Since $\Pr_\U\bigl( (q,\epsilon) \models \Diamond (x,\epsilon)_u
\bigr) \gr 0$ this yields
\[
\Pr\nolimits_\U\bigl((q,\epsilon) \models \bigvee\limits_{i=1}^n
(\Box \Diamond A_i \wedge \Diamond \Box B_i)\bigr) \ls 1,
\]
which contradicts assumption (1).  We conclude
$\Pr_\U\bigl((q,\epsilon) \models \Diamond C\bigr)= 1$.

(2) $\Longrightarrow$ (1): Let $\V$ be as in (2).  We may assume that
$\V$ is memoryless (see Lemmas~\ref{lem-prom2} and \ref{lem-prom1}).
For any location $z \in C$, we choose a finite-memory scheduler
$\V_z$ for $\N$ such that
\[
\Pr\nolimits_{\V_z}\bigl((z,\epsilon) \models (\Box \Diamond A_i
\wedge \Diamond \Box B_i)\bigr) = 1
\]
for some $i \in \{1,\ldots,n\}$.  Let $\U$ be the finite-memory
scheduler that first behaves as $\V$, reaching $C$ almost surely, and
which, after having visited a location $z \in C$, mimics the
schedulers $\V_z$ as follows.  When entering $C$ the first time, say
in configuration $(z,w)$ where $w \neq \varepsilon$, then $\U$ goes
into a waiting mode where it waits until a configuration
$(z',\epsilon)$ with $z' \in C$ has been entered.  From this
configuration $(z',\epsilon)$ on, $\U$ behaves as $\V_{z'}$.  In the
waiting mode, $\U$ chooses the same transition rule for $(z,w)$ as
$\V_z$ for the starting configuration $(z,\epsilon)$.

Note that the configurations obtained from $(z,w)$ by taking this
transition rule have the form $(z',w')$ where $z' \in C$.  This is
because $(z',\epsilon)$ is a successor of $(z,\epsilon)$ under
this transition rule. Hence, $\V_z$ induces a scheduler under which
$(z',\epsilon)$ fulfills $\Box \Diamond A_i \wedge \Diamond \Box
B_i$ almost surely for some index $i$.  This yields $z' \in C_i
\subseteq C$.

The finite attractor property yields that $\U$ will eventually leave
the waiting mode. Thus, $\U$ has the desired property.


\section{$\omega$-regular properties}
\label{sec-omegareg}

We now consider qualitative verification of $\omega$-regular linear-time
properties where, as before, we use 
the control locations of the underlying
NPLCS as atomic propositions (with the obvious interpretation). 

For algorithmic purposes, we assume that an 
$\omega$-regular property is given
by a deterministic (word) Streett automaton with the alphabet
$Q$ (the set of control locations in the given NPLCS).
Other equivalent formalisms (nondeterministic Streett automata,
nondeterministic B\"uchi automata, $\mu$-calculus formulas, etc.) are of
course possible. The translations 
between them is now well understood. 
See, e.g., the survey articles in \cite{Graedel-Thomas-Wilke-LNCS}.
\\

A deterministic Streett automaton
(DSA for short) over the alphabet $Q$ 
is a tuple 
$\A =(Z,\sigma,z_0,\Acc)$ where $Z$ is a finite set of states, 
$\sigma:Z\times Q\to Z$ the transition function, 
$z_0 \in Z$ the initial state,
and $\Acc = \{ (A_1,B_1),\ldots, (A_n,B_n)\}$ 
a set of pairs $(A_i,B_i)$ consisting of subsets $A_i,B_i \subseteq Z$.
$\Acc$ is called the acceptance condition of $\A$. 
Intuitively, $\Acc$ 
stands for the strong fairness condition
$\psi_\A = \bigwedge_{i=1}^{n} 
(\Box \Diamond A_i \Rightarrow \Box \Diamond B_i)$.
The accepting language $L(\A)$ consists of all infinite words
$q_0,q_1,q_2,\ldots \in Q^\omega$ 
where the induced run $z_0 \step{q_0} z_1 \step{q_1} z_2 \step{q_2} 
\cdots$ in $\A$
(which is obtained by starting in the initial state $z_0$ of $\A$
and putting $z_{j+1} = \sigma(z_j,q_{j})$, $j=0,1,2,\ldots$)
is accepting, that is,
for all $i \in \{1,\ldots,n\}$,
$z_j \in A_i$ for at most finitely many indices $j$ or
$z_j \in B_i$ for infinitely many indices $j$.
For a path $\pi$ of some NPLCS with state space $Q$, 
we write $\pi\sat\A$ when $\pi$ (more precisely, its
projection over $Q^\omega$) belongs to $L(\A)$.

Since Streett properties are $\omega$-regular, 
Theorem~\ref{thm:Streett-undec} immediately
entails:
\begin{corollary}[($\omega$-regular properties)]
The problem, given NPLCS $\N$, location $q\in Q$, and DSA $\A$, whether
there exists a scheduler $\U$ with $\Pr_\U\bigl((q,\epsilon)\models
\A\bigr) $= 1 (or $\ls 1$, or $=0$, or $\gr 0$), is undecidable.
\end{corollary}

More interesting is the fact that our positive results from
section~\ref{sec-finite-mem} carry over from Streett properties to all
$\omega$-regular properties:

\begin{theorem}[($\omega$-regular properties, finite-memory schedulers)] 
\label{thm:omega-regular-fm-schedulers}
  The problem, given NPLCS $\N$, location $q\in Q$, and DSA $\A$, whether
  there exists a \emph{finite-memory} scheduler 
  $\U$ such that $\Pr_\U\bigl((q,\epsilon)\models \A\bigr)=1$ (or
  $\ls 1$, or $=0$, or $\gr 0$), is decidable.
\end{theorem}

The extension from repeated-reachability properties to $\omega$-regular
properties follows the standard automata-theoretic approach for the
verification of qualitative properties: one reduces the question whether
$\N$ is accepted by $\A$ to a repeated-reachability property over the
``product'' $\N \times \A$ (see, e.g.,~\cite{Vardi99}).
We briefly sketch the main steps of the reduction which yields the
proof for Theorem \ref{thm:omega-regular-fm-schedulers}.

Let $\N$ be a NPLCS and $\A$ a DSA as before. The product $\N' \egdef
\N \times \A$ is a NPLCS where:
\begin{itemize}
\item locations are pairs $(p,z)$ where $p \in Q$ is a location in
  $\N$ and $z \in Z$ a state of $\A$,
\item the channel set and the message alphabet are as in  $\N$,
\item $(p,z) \step{\op} (r,z')$ is a transition rule in $\N
  \times \A$ if and only if $p \step{\op} r$ is a transition rule in
  $\N$ and $z'=\sigma(z,p)$. 
\end{itemize}
Then, each infinite path $\pi$ in $\N$, of the general form
\begin{gather}
\tag{$\pi$}
           (q_0,w_0)\to (q_1,w_1) \to (q_2,w_2) \to (q_3,w_3) \cdots
\end{gather}
is lifted to a path $\pi'$ in $\N \times \A$
\begin{gather}
\tag{$\pi'$}
(q_0,z_0,w_0) \to 
  (q_1,z_1,w_1) \to (q_2,z_2,w_2) \to (q_3,z_3,w_3) \cdots
\end{gather}
where $z_{j+1}\egdef \sigma(z_j,q_j)$ for all $j\in\Nat$. Thus,
$z_0\step{q_0}z_1\step{q_1}z_2\step{q_2}\cdots$ is the (unique) 
run of $\A$
on (the projection of) $\pi$. 
Vice versa, any path $\pi'$ in $\N \times \A$
arises through the combination of a path in $\N$ and its run in $\A$.

Assume the acceptance condition of $\A$ is given by the following
Streett condition: $\psi_\A = \bigwedge_{i=1}^n (\Box \Diamond
A_i\impl \Box \Diamond B_i)$ with $A_i,B_i\subseteq Z$. Then, letting
$A'_i \egdef Q \times A_i$ and $B'_i \egdef Q \times B_i$, we equip
$\N \times \A$ with the acceptance condition $\Acc' = \{ (A_i',B_i') :
1 \leq i \leq n\}$ which corresponds to the following Streett
condition $\psi_{\N \times \A}$:
\begin{gather}
\tag{$\psi_{\N \times \A}$}
\bigwedge\limits_{i=1}^n (\Box \Diamond A'_i \impl \Box \Diamond
B'_i)
\end{gather}

\begin{lemma}
  Let $\pi$ be a path in $\N$ and $\pi'$ the corresponding path in
  $\N'$. Then, $\pi \sat \A$ if and only if 
$\pi' \models \psi_{\N \times \A}$.
\end{lemma}

This correspondence between paths in $\N$ and paths in $\N'$ allows to
transform any scheduler $\U$ for $\N$ into a scheduler $\V$ for $\N'$
such that the probability agrees and vice versa.
More precisely:

\begin{lemma}
\label{lemma-A2psi}
Let $p\in [0,1]$, then there exists a finite-memory scheduler $\U$ 
for $\N$
such that 
$\Pr_\U\bigl((q,\epsilon)\models \A\bigr) =p$ iff there exists a
finite-memory scheduler $\V$ for $\N \times \A$ 
s.t.\
$$\Pr\nolimits_\V\bigl((q,z_0,\epsilon)\models \psi_{\N \times \A})
=p.
$$
\end{lemma}
The proof is as in \cite[section~4]{Courcoubetis95}, the basic ingredient
being that $\A$ is deterministic.


Lemma~\ref{lemma-A2psi} reduces  the verification of qualitative
$\omega$-regular properties over $\N$ to the verification of
qualitative Streett properties over $\N'$.
Decidability is then obtained with Theorem~\ref{thm:streett-fm}.


\section{Conclusion}
\label{sec-conclusion}

We proposed NPLCS's, a model for lossy channel systems where message losses
occur probabilistically while transition rules behave nondeterministically,
and we investigated qualitative verification problems for this model. Our
main result is that qualitative verification of simple linear-time properties
is decidable, but this does not extend to all $\omega$-regular properties.
On the other hand, decidability is recovered if we restrict our attention
to finite-memory schedulers.

The NPLCS model improves on earlier models for lossy channel systems: the
original, purely nondeterministic, LCS model is too pessimistic w.r.t.\
message losses and nondeterministic losses make liveness properties
undecidable. It seems this undecidability is an artifact of the standard
rigid view asking whether no incorrect behavior exists, when we could be
content with the weaker statement that incorrect behaviors are extremely
unlikely. The fully probabilistic PLCS model recovers decidability but
cannot account for nondeterminism.

Regarding NPLCS's, decidability is obtained by reducing qualitative
properties to reachability questions in the underlying non-probabilistic
transition system. Since in our model qualitative properties do not depend
on the exact value of the fault rate $\tau$, the issue of what is a
realistic value for $\tau$ is avoided, and one can establish correctness
results that apply uniformly to all fault rates.

An important open question is the decidability of quantitative properties.
Regarding this research direction, we note that \citeN{rabinovich2003}
investigated it for the fully probabilistic PLCS model, where it already
raises serious difficulties.



\bibliographystyle{acmtrans}

\begin{thebibliography}{}

\bibitem[\protect\citeauthoryear{Abdulla, Baier, {Purushothaman Iyer}, and
  Jonsson}{Abdulla et~al\mbox{.}}{2005}]{abdulla2005}
{\sc Abdulla, P.~A.}, {\sc Baier, C.}, {\sc {Purushothaman Iyer}, S.}, {\sc
  and} {\sc Jonsson, B.} 2005.
\newblock Simulating perfect channels with probabilistic lossy channels.
\newblock {\em Information and Computation\/}~{\em 197,\/}~1--2, 22--40.

\bibitem[\protect\citeauthoryear{Abdulla, Bertrand, Rabinovich, and
  Schnoebelen}{Abdulla et~al\mbox{.}}{2005}]{ABRS-icomp}
{\sc Abdulla, P.~A.}, {\sc Bertrand, N.}, {\sc Rabinovich, A.}, {\sc and} {\sc
  Schnoebelen, {\relax Ph}.} 2005.
\newblock Verification of probabilistic systems with faulty communication.
\newblock {\em Information and Computation\/}~{\em 202,\/}~2, 141--165.

\bibitem[\protect\citeauthoryear{Abdulla, {\v{c}}er{\=a}ns, Jonsson, and
  Tsay}{Abdulla et~al\mbox{.}}{2000}]{abdulla2000c}
{\sc Abdulla, P.~A.}, {\sc {\v{c}}er{\=a}ns, K.}, {\sc Jonsson, B.}, {\sc and}
  {\sc Tsay, Y.-K.} 2000.
\newblock Algorithmic analysis of programs with well quasi-ordered domains.
\newblock {\em Information and Computation\/}~{\em 160,\/}~1/2, 109--127.

\bibitem[\protect\citeauthoryear{Abdulla, {Collomb-Annichini}, Bouajjani, and
  Jonsson}{Abdulla et~al\mbox{.}}{2004}]{abdulla-forward-lcs}
{\sc Abdulla, P.~A.}, {\sc {Collomb-Annichini}, A.}, {\sc Bouajjani, A.}, {\sc
  and} {\sc Jonsson, B.} 2004.
\newblock Using forward reachability analysis for verification of lossy channel
  systems.
\newblock {\em Formal Methods in System Design\/}~{\em 25,\/}~1, 39--65.

\bibitem[\protect\citeauthoryear{Abdulla and Jonsson}{Abdulla and
  Jonsson}{1996a}]{abdulla96c}
{\sc Abdulla, P.~A.} {\sc and} {\sc Jonsson, B.} 1996a.
\newblock Undecidable verification problems for programs with unreliable
  channels.
\newblock {\em Information and Computation\/}~{\em 130,\/}~1, 71--90.

\bibitem[\protect\citeauthoryear{Abdulla and Jonsson}{Abdulla and
  Jonsson}{1996b}]{abdulla96b}
{\sc Abdulla, P.~A.} {\sc and} {\sc Jonsson, B.} 1996b.
\newblock Verifying programs with unreliable channels.
\newblock {\em Information and Computation\/}~{\em 127,\/}~2, 91--101.

\bibitem[\protect\citeauthoryear{Baier, Bertrand, and Schnoebelen}{Baier
  et~al\mbox{.}}{2006}]{BBS-attractor}
{\sc Baier, C.}, {\sc Bertrand, N.}, {\sc and} {\sc Schnoebelen, {\relax Ph}.}
  2006.
\newblock A note on the attractor-property of infinite-state {Markov} chains.
\newblock {\em Information Processing Letters\/}~{\em 97,\/}~2, 58--63.

\bibitem[\protect\citeauthoryear{Baier and Engelen}{Baier and
  Engelen}{1999}]{baier99}
{\sc Baier, C.} {\sc and} {\sc Engelen, B.} 1999.
\newblock Establishing qualitative properties for probabilistic lossy channel
  systems: An algorithmic approach.
\newblock In {\em Proc.\ 5th Int.\ AMAST Workshop Formal Methods for Real-Time
  and Probabilistic Systems (ARTS '99), Bamberg, Germany, May 1999}. Lecture
  Notes in Computer Science, vol. 1601. Springer, 34--52.

\bibitem[\protect\citeauthoryear{Bertrand and Schnoebelen}{Bertrand and
  Schnoebelen}{2003}]{BS03}
{\sc Bertrand, N.} {\sc and} {\sc Schnoebelen, {\relax Ph}.} 2003.
\newblock Model checking lossy channels systems is probably decidable.
\newblock In {\em Proc.\ 6th Int.\ Conf.\ Foundations of Software Science and
  Computation Structures (FOSSACS 2003), Warsaw, Poland, Apr.\ 2003}. Lecture
  Notes in Computer Science, vol. 2620. Springer, 120--135.

\bibitem[\protect\citeauthoryear{Bertrand and Schnoebelen}{Bertrand and
  Schnoebelen}{2004}]{BS04}
{\sc Bertrand, N.} {\sc and} {\sc Schnoebelen, {\relax Ph}.} 2004.
\newblock Verifying nondeterministic channel systems with probabilistic message
  losses.
\newblock In {\em Proc.\ 3rd Int.\ Workshop on Automated Verification of
  Infinite-State Systems (AVIS 2004), Barcelona, Spain, Apr.\ 2004},
  {R.~Bharadwaj}, Ed.

\bibitem[\protect\citeauthoryear{Brand and Zafiropulo}{Brand and
  Zafiropulo}{1983}]{brand83}
{\sc Brand, D.} {\sc and} {\sc Zafiropulo, P.} 1983.
\newblock On communicating finite-state machines.
\newblock {\em Journal of the ACM\/}~{\em 30,\/}~2, 323--342.

\bibitem[\protect\citeauthoryear{C\'ec\'e, Finkel, and {Purushothaman
  Iyer}}{C\'ec\'e et~al\mbox{.}}{1996}]{cece95}
{\sc C\'ec\'e, G.}, {\sc Finkel, A.}, {\sc and} {\sc {Purushothaman Iyer}, S.}
  1996.
\newblock Unreliable channels are easier to verify than perfect channels.
\newblock {\em Information and Computation\/}~{\em 124,\/}~1, 20--31.

\bibitem[\protect\citeauthoryear{Courcoubetis and Yannakakis}{Courcoubetis and
  Yannakakis}{1995}]{Courcoubetis95}
{\sc Courcoubetis, C.} {\sc and} {\sc Yannakakis, M.} 1995.
\newblock The complexity of probabilistic verification.
\newblock {\em Journal of the ACM\/}~{\em 42,\/}~4, 857--907.

\bibitem[\protect\citeauthoryear{Emerson}{Emerson}{1990}]{Emerson-Handbook}
{\sc Emerson, E.~A.} 1990.
\newblock Temporal and modal logic.
\newblock In {\em Handbook of Theoretical Computer Science}, {J.~v. Leeuwen},
  Ed. Vol.~B. Elsevier Science, Chapter~16, 995--1072.

\bibitem[\protect\citeauthoryear{Finkel}{Finkel}{1994}]{finkel94}
{\sc Finkel, A.} 1994.
\newblock Decidability of the termination problem for completely specificied
  protocols.
\newblock {\em Distributed Computing\/}~{\em 7,\/}~3, 129--135.

\bibitem[\protect\citeauthoryear{Finkel and Schnoebelen}{Finkel and
  Schnoebelen}{2001}]{finkel-wsts}
{\sc Finkel, A.} {\sc and} {\sc Schnoebelen, {\relax Ph}.} 2001.
\newblock Well-structured transition systems everywhere!
\newblock {\em Theoretical Computer Science\/}~{\em 256,\/}~1--2, 63--92.

\bibitem[\protect\citeauthoryear{Gr{\"{a}}del, Thomas, and Wilke}{Gr{\"{a}}del
  et~al\mbox{.}}{2002}]{Graedel-Thomas-Wilke-LNCS}
{\sc Gr{\"{a}}del, E.}, {\sc Thomas, W.}, {\sc and} {\sc Wilke, T.}, Eds. 2002.
\newblock {\em Automata, Logics, and Infinite Games: A Guide to Current
  Research}. Lecture Notes in Computer Science, vol. 2500.
\newblock Springer.

\bibitem[\protect\citeauthoryear{Kemeny, Snell, and Knapp}{Kemeny
  et~al\mbox{.}}{1966}]{Kemeny-Snell}
{\sc Kemeny, J.~G.}, {\sc Snell, J.~L.}, {\sc and} {\sc Knapp, A.~W.} 1966.
\newblock {\em Denumerable {Markov} Chains}.
\newblock D.\ Van Nostrand Co., Princeton, NJ, USA.

\bibitem[\protect\citeauthoryear{Masson and Schnoebelen}{Masson and
  Schnoebelen}{2002}]{MS-mfcs2002}
{\sc Masson, B.} {\sc and} {\sc Schnoebelen, {\relax Ph}.} 2002.
\newblock On verifying fair lossy channel systems.
\newblock In {\em Proc.\ 27th Int.\ Symp.\ Math.\ Found.\ Comp.\ Sci.\ (MFCS
  2002), Warsaw, Poland, Aug.\ 2002}. Lecture Notes in Computer Science, vol.
  2420. Springer, 543--555.

\bibitem[\protect\citeauthoryear{Mayr}{Mayr}{2003}]{Mayr-unreliable}
{\sc Mayr, R.} 2003.
\newblock Undecidable problems in unreliable computations.
\newblock {\em Theoretical Computer Science\/}~{\em 297,\/}~1--3, 337--354.

\bibitem[\protect\citeauthoryear{Pachl}{Pachl}{1987}]{pachl87}
{\sc Pachl, J.~K.} 1987.
\newblock Protocol description and analysis based on a state transition model
  with channel expressions.
\newblock In {\em Proc.\ 7th IFIP WG6.1 Int.\ Workshop on Protocol
  Specification, Testing, and Verification (PSTV '87), Zurich, Switzerland, May
  1987}. North-Holland, 207--219.

\bibitem[\protect\citeauthoryear{Panangaden}{Panangaden}{2001}]{panangaden2001}
{\sc Panangaden, P.} 2001.
\newblock Measure and probability for concurrency theorists.
\newblock {\em Theoretical Computer Science\/}~{\em 253,\/}~2, 287--309.

\bibitem[\protect\citeauthoryear{{Purushothaman Iyer} and
  Narasimha}{{Purushothaman Iyer} and Narasimha}{1997}]{purush97}
{\sc {Purushothaman Iyer}, S.} {\sc and} {\sc Narasimha, M.} 1997.
\newblock Probabilistic lossy channel systems.
\newblock In {\em Proc.\ 7th Int.\ Joint Conf.\ Theory and Practice of Software
  Development (TAPSOFT '97), Lille, France, Apr.\ 1997}. Lecture Notes in
  Computer Science, vol. 1214. Springer, 667--681.

\bibitem[\protect\citeauthoryear{Puterman}{Puterman}{1994}]{Puterman94}
{\sc Puterman, M.~L.} 1994.
\newblock {\em {Markov} decision processes: discrete stochastic dynamic
  programming}.
\newblock John Wiley \& Sons.

\bibitem[\protect\citeauthoryear{Rabinovich}{Rabinovich}{2003}]{rabinovich2003}
{\sc Rabinovich, A.} 2003.
\newblock Quantitative analysis of probabilistic lossy channel systems.
\newblock In {\em Proc.\ 30th Int.\ Coll.\ Automata, Languages, and Programming
  (ICALP 2003), Eindhoven, NL, July 2003}. Lecture Notes in Computer Science,
  vol. 2719. Springer, 1008--1021.

\bibitem[\protect\citeauthoryear{Schnoebelen}{Schnoebelen}{2001}]{Sch-tacs2001}
{\sc Schnoebelen, {\relax Ph}.} 2001.
\newblock Bisimulation and other undecidable equivalences for lossy channel
  systems.
\newblock In {\em Proc.\ 4th Int.\ Symp.\ Theoretical Aspects of Computer
  Software (TACS 2001), Sendai, Japan, Oct.\ 2001}. Lecture Notes in Computer
  Science, vol. 2215. Springer, 385--399.

\bibitem[\protect\citeauthoryear{Schnoebelen}{Schnoebelen}{2002}]{phs-IPL2002}
{\sc Schnoebelen, {\relax Ph}.} 2002.
\newblock Verifying lossy channel systems has nonprimitive recursive
  complexity.
\newblock {\em Information Processing Letters\/}~{\em 83,\/}~5, 251--261.

\bibitem[\protect\citeauthoryear{Schnoebelen}{Schnoebelen}{2004}]{Sch-voss}
{\sc Schnoebelen, {\relax Ph}.} 2004.
\newblock The verification of probabilistic lossy channel systems.
\newblock In {\em Validation of Stochastic Systems -- A Guide to Current
  Research}, {C.~Baier} {et~al\mbox{.}}, Eds. Lecture Notes in Computer
  Science, vol. 2925. Springer, 445--465.

\bibitem[\protect\citeauthoryear{Vardi}{Vardi}{1985}]{Vardi85}
{\sc Vardi, M.~Y.} 1985.
\newblock Automatic verification of probabilistic concurrent finite-state
  programs.
\newblock In {\em Proc.\ 26th Symp.\ Foundations of Computer Science (FOCS
  '85), Portland, OR, USA, Oct.\ 1985}. IEEE Comp.\ Soc.\ Press, 327--338.

\bibitem[\protect\citeauthoryear{Vardi}{Vardi}{1999}]{Vardi99}
{\sc Vardi, M.~Y.} 1999.
\newblock Probabilistic linear-time model checking: An overview of the
  automata-theoretic approach.
\newblock In {\em Proc.\ 5th Int.\ AMAST Workshop Formal Methods for Real-Time
  and Probabilistic Systems (ARTS '99), Bamberg, Germany, May 1999}. Lecture
  Notes in Computer Science, vol. 1601. Springer, 265--276.

\end{thebibliography}


\newpage

\appendix

\section{Proof of Lemma~\ref{lem-prom-union}}
\label{appendix-prom}

The goal is to prove that given $A$ and $B$ sets of locations, $\Prom(A
\cup B) = \Prom(A) \cup \Prom(B)$. One inclusion is trivial: $\Prom(A)
\cup \Prom(B) \subseteq \Prom(A \cup B)$. We prove here the reverse
inclusion. In fact we build a scheduler that, starting from any
$(x,\epsilon)$ for $x \in \Prom(A \cup B)$, will ensure visiting
eventually $A$ or visiting eventually $B$, and the choice between $A$
and $B$ is fixed (given $x$).  Lemma~\ref{lem-prom2} then yields
$\Prom(A \cup B) \subseteq \Prom(A) \cup \Prom(B)$.

For each $x \in \Prom(A \cup B)$ we pick a simple path to $A \cup B$,
that only visits locations of $\Prom(A \cup B)$.  Such a path exists
by definition of $\Prom$, we denote it
\[
\pi_x :\ (x,\epsilon)=(x^0,w_x^0) \step{\delta_x^0} (x^1,w_x^1)
\step{\delta_x^1} (x^2,w_x^2) \cdots \step{\delta_x^{m-1}} (x^m,w_x^m)
\]
with $x^m \in A \cup B$ and $x^i \in \Prom(A \cup B)$ for $i \ls m$. By
convention, we let $x^i=x^{|\pi_x|}$ when $i \gr |\pi_x|$.
\begin{figure}[htbp]
\centering
\includegraphics{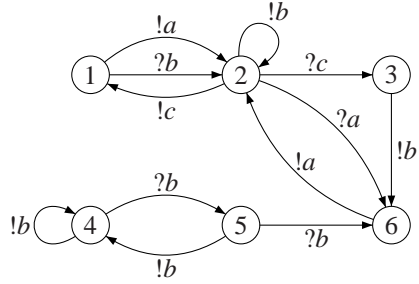}
\caption{Running example for the proof of Lemma~\ref{lem-prom-union}}
\label{fig-ex-prom}
\end{figure}
For example, given the system depicted in Fig.~\ref{fig-ex-prom}, with
$A=\{3\}$ and $B=\{6\}$, one has $\Prom(A \cup B)= \{1,2,3,4,5,6\}$
and a possible choice for the paths $\pi_x$ is given in
Fig.~\ref{table-pi-f1}.


We now define a sequence $\Pcal_0, \Pcal_1, \ldots$ of
partitions of $\Prom(A \cup B)$. In general $\Pcal_k$ is some
$\{B_1^k,B_2^k,\ldots\}$ and each class $B_j^k \in \Pcal_k$
comes with a fixed element $b_j^k$ called its \emph{representative}
(which is underlined in the examples). The first partition is composed
of all singletons: $\Pcal_0=\{\{x\} \mid x \in \Prom(A \cup
B)\}$.  Partition $\Pcal_{k+1}$ is coarser than $\Pcal_k$:
each class in $\Pcal_{k+1}$ is the fusion of (possibly only one)
classes of $\Pcal_k$. Assume $\Pcal_k$ is given:
$\Pcal_k =\{B_1^k, \ldots \}$ with $\{b_1^k, \ldots \}$ as
representatives. We define a mapping $f_{k+1}$ between the classes of
$\Pcal_k$. For any class $B_j^k$, we consider its representative
$b_j^k$, shortly written $x$, and associate with $B_j^k$ the class to
which $x^{k+1}$ (the $k+1$-th location on $\pi_x$) belongs. In our
running example $f_1$ is given on Fig.~\ref{table-pi-f1}.
\begin{figure}[htbp]
\newlength{\offsetleng}
\newlength{\spreadleng}
\setlength{\offsetleng}{-2ex}
\setlength{\spreadleng}{3ex}
\vspace{-1em}
\centering
\begin{tabular}{r>{\hspace{\offsetleng}}c<{\hspace{\offsetleng}}l 
                c<{\hspace{\spreadleng}} 
                r>{\hspace{\offsetleng}}c<{\hspace{\offsetleng}}l 
                c<{\hspace{\spreadleng}}
                r>{\hspace{\offsetleng}}c<{\hspace{\offsetleng}}l
                }
    $\pi_1$ & $\egdef$ & $(1,\epsilon) \to (2,a) \to (6,\epsilon)$
& & $f_1(\{\underline{1}\})$ & $=$ & $\{2\}$
& & & &
\\
    $\pi_2$ & $\egdef$ & $(2,\epsilon) \to (2,b) \to (1,bc) \to (2,c) \to (3,\epsilon)$
& & $f_1(\{\underline{2}\})$ & $=$ & $\{2\}$ 
& & $f_4(\{1,\underline{2}\})$ & $=$ & $\{3\}$
\\
    $\pi_3$ & $\egdef$ & $(3,\epsilon)$ 
& & $f_1(\{\underline{3}\})$ & $=$ & $\{3\}$ 
& & $f_4(\{\underline{3}\})$ & $=$ & $\{3\}$
\\
    $\pi_4$ & $\egdef$ & $(4,\epsilon)\to (4,b)\to (4,bb)\to (5,b)\to (6,\epsilon)$ 
& & $f_1(\{\underline{4}\})$ & $=$ & $\{4\}$ 
& & $f_4(\{\underline{4},5\})$ & $=$ & $\{6\}$
\\
    $\pi_5$ & $\egdef$ & $(5,\epsilon)\to (4,b)\to (4,bb)\to (5,b)\to (6,\epsilon)$ 
& & $f_1(\{\underline{5}\})$ & $=$ & $\{4\}$ 
& & & &
\\
    $\pi_6$ & $\egdef$ & $(6,\epsilon)$ 
& & $f_1(\{\underline{6}\})$ & $=$ & $\{6\}$
& & $f_4(\{\underline{6}\})$ & $=$ & $\{6\}$ 
\end{tabular}
\caption{Running example (continued): paths $\pi_x$ with mappings $f_1$ and $f_4$}
\label{table-pi-f1}
\end{figure}

The mapping $f_{k+1}$ induces an oriented graph (of out-degree $1$).
The classes of $\Pcal_{k+1}$ are obtained by fusing the classes
of $\Pcal_k$ which belong to the same connected component in
this graph.
\begin{figure}[htbp]
\centering
\includegraphics{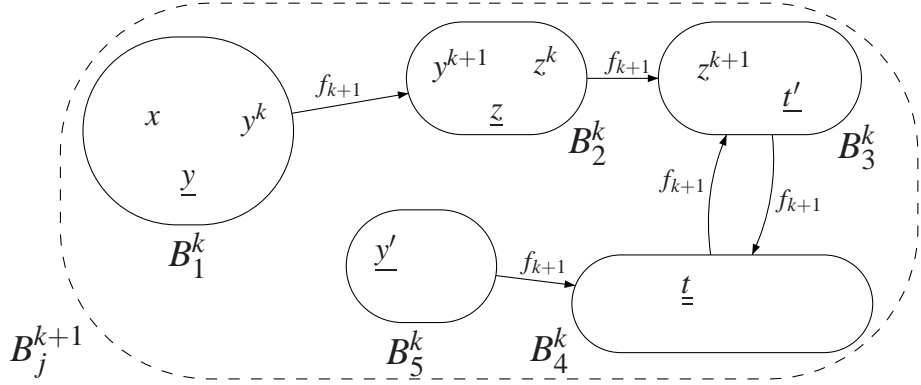}
\caption{Constructing $\Pcal_{k+1}$ by fusing equivalence classes from $\Pcal_k$}
\label{fig-proof-appendix}
\end{figure}
For example, in Fig.~\ref{fig-proof-appendix}, the classes $B_1^k$,
$B_2^k$, $B_3^k$, $B_4^k$ and $B_5^k$ are fused. The representative
for $B^{k+1}_j$ is arbitrarily chosen among the representatives of the
$B^k_i$'s that compose the strongly connected component ($b_3^k$ or
$b_4^k$ in Fig.~\ref{fig-proof-appendix}). Back to the running
example, we derive $\Pcal_1 =
\{\{1,\underline{2}\},\{\underline{3}\},\{\underline{4},5\},\{\underline{6}\}
\}$ with ${2,3,4,6}$ as representatives (no choice here). Partition
$\Pcal_1$ is stable by $f_2$ and $f_3$.
\begin{gather*}
  f_2, f_3 :\ \{1,\underline{2}\} \to \{1,2\} \ \ \ \ 
  \{\underline{3}\} \to \{3\}\ \ \ \ \{\underline{4},5\} \to \{4,5\}\ 
  \ \ \ \{\underline{6}\} \to \{6\}
\end{gather*}
Hence $\Pcal_3=\Pcal_2=\Pcal_1$. Mapping $f_4$ is given in
Fig.~\ref{table-pi-f1}. We deduce
$\Pcal_4=\{\{1,2,\underline{3}\},\{4,5,\underline{6}\} \}$ with $3$
and $6$ as representatives (no choice either).

It is clear that $\Pcal_{k+1}$ is coarser than $\Pcal_k$
and that a representative at level $k+1$ was already a representative
at level $k$. Hence the sequence eventually stabilizes (the state
space is finite). We denote $\Pcal_\infty =\{B^\infty_1,
\ldots\}$ the partition in the limit. In the running example
$\Pcal_\infty = \Pcal_4$.

This whole construction is geared towards the following:
\begin{lemma}
\label{lem-prom-crucial}
For all $k \geq 1$, there exists a scheduler $\U_k$ such that, for
every class $B_j^k$, and writing $y$ for $b_j^k$,
\begin{gather}
\tag{*}
\label{sched-crucial}
\forall x \in B_j^k \;\; \forall w \in {\Mess^*}^\C \;\; 
\Pr\nolimits_{\U_k}\bigl( (x,w) \models \Diamond (y^k,w_y^k) \bigr) =1
\end{gather}
\end{lemma}

In other words, at step $k$ of the construction there exists a
scheduler that, starting from a location $x$ with arbitrary channel
content, ensures (with probability one) we'll visit the $k$-th
configuration on $\pi_y$ where $y$ is the representative for $x$ in
$\Pcal_k$. When $k$ is large enough, more precisely larger than
all $|\pi_x|$'s, \eqref{sched-crucial} states that $\U_k$ guarantees
reaching $A$ (or $B$, depending on $x$) with probability one, which
concludes the proof of Lemma~\ref{lem-prom-union}.
\begin{proof}[(of Lemma~\ref{lem-prom-crucial})]
The proof is by induction on $k$.

We first prove the case $k=1$. Let $x$ be a location in some class
$B_i^1$ having $(y=)b_i^1$ as representative. The behavior of $\U_1$
is simple: in any configuration $(z,v)$, $\U_1$ fires $\delta_z^0$.
Going on this way, $\U_1$ eventually ends up in the strongly connected
component (w.r.t $f_1$).  Because of the finite-attractor property, the
configuration $(y,\epsilon)$ is visited infinitely often almost
surely. Hence, $\U_1$ will succeed in reaching $(y^1,w_y^1)$ from
$(y,\epsilon)$ by $\delta_y^1$.

Assume now that for some $k \geq 1$ there exists $\U_k$ ensuring
\eqref{sched-crucial}. We consider $\Pcal_{k+1}$ and build
$\U_{k+1}$, using $\U_k$. Let $x \in B_j^{k+1}$ (it may help to look
at Fig.~\ref{fig-proof-appendix}). Starting from $(x,w)$, $\U_{k+1}$
behaves as $\U_k$ until $(y^k,w_y^k)$ is reached. Then it fires
$\delta_y^k$ and ends up in $(y^{k+1},w')$ for some channel content
$w'$. $y^{k+1}$ is a location of $B_{i'}^k=f_{k+1}(B_i^k)$; let
$z=b_{i'}^k$ be its representative. From configuration $(y^{k+1},w')$,
$\U_{k+1}$ behaves again as $\U_k$ and eventually reaches
$(z^k,w_z^k)$ with probability one.  Iterating this process
(alternation of $\U_k$'s behavior and one step transition), $\U_{k+1}$
will eventually end in the strongly connected component of
$B_j^{k+1}$. If $t$ is the representative for this class in
$\Pcal_{k+1}$, because of the finite-attractor property
$(t,\epsilon)$ is visited infinitely often, almost surely. Hence,
$\U_{k+1}$ will in the end succeed and reach $(t^{k+1},w_t^{k+1})$
using $\U_k$ until $(t^k,w_t^k)$ and then performing $\delta_t^k$.
\end{proof}



\begin{received}
Received November 2005;
accepted March 2006
\end{received}
\end{document}